\documentclass[prb, twocolumn, superscriptaddress]{revtex4-2}
\usepackage{bm, amsmath, amsfonts, amssymb, mathtools, ascmac, braket}
\usepackage{multirow}
\usepackage{graphicx}
\usepackage{float, xcolor}
\usepackage{physics}
\usepackage{comment}

\usepackage[
pagebackref=false,
colorlinks=true,
linkcolor=blue,
urlcolor=blue,
filecolor=black,
citecolor=blue,
pdfstartview=FitV,
pdftitle={},
pdfauthor={},
pdfsubject={},
pdfkeywords={},
pdfpagemode=None,
bookmarksopen=true
]{hyperref}

\newcommand{\bk}{\mathbf{k}}
\newcommand{\bn}{\mathbf{n}}
\newcommand{\BR}{\mathbf{R}}
\newcommand{\T}{\mathbb{T}}
\newcommand{\bd}{\mathbf{d}}
\DeclareMathOperator{\MH}{\mathcal{H}}

\DeclareMathOperator{\sgn}{\text{sgn}}

\begin{document}

\title{Abelian spectral topology of multifold exceptional points}
\author{Marcus St\aa lhammar}
\email{marcus.backlund@physics.uu.se}
\thanks{Acts as first, senior and corresponding author.}
\affiliation{Institute for Theoretical Physics, Utrecht University, Princetonplein 5, 3584CC Utrecht, The Netherlands} \affiliation{Department of Physics and Astronomy, Uppsala University, Uppsala, Sweden}
\author{Lukas R\o dland}
\affiliation{Department of Physics, Stockholm University, AlbaNova University Center, 10691 Stockholm, Sweden}
\date{\today}

\begin{abstract}
    The advent of non-Hermitian physics has enriched the plethora of topological phases to include phenomena without Hermitian counterparts. 
    Despite being among the most well-studied uniquely non-Hermitian features, the topological properties of multifold exceptional points, $n$-fold spectral degeneracies (EP$n$s) at which also the corresponding eigenvectors coalesce, were only recently revealed in terms of topological resultant winding numbers and concomitant Abelian doubling theorems. 
    Nevertheless, a more mathematically fundamental description of EP$n$s and their topological nature has remained an open question. 
    To fill this void, in this article, we revisit the topological classification of EP$n$s in generic systems and systems with local symmetries, generalize it in terms of more mathematically tractable (local) similarity relations, and extend it to include all such similarities as well as non-local symmetries.
    Through the resultant vector, whose components are given in terms of the resultants between the corresponding characteristic polynomial and its derivatives, the topological nature of the resultant winding number is understood in several ways: in terms of i) the tenfold classification of (Hermitian) topological matter, ii) the framework of Mayer--Vietoris sequence, and iii) the classification of vector bundles.
    The classification scheme further predicts the existence of topological bulk Fermi arcs protected by a $\mathbb{Z}_2$-invariant, induced by non-local symmetries, dubbed $\mathbb{Z}_2$-protected Fermi arcs.
    Our work reveals the mathematical foundations on which the topological nature of EP$n$s resides, enriches the theoretical understanding of non-Hermitian spectral features, and will therefore find great use in modern experiments within both classical and quantum physics. 
\end{abstract}

\maketitle

\section{Introduction}
The mathematical branch of topology has comprised an important asset in theoretical physics during the last century, with applications ranging from high-energy physics and topological quantum field theory~\cite{TQFTBook}, to modern photonics~\cite{Topphot,Lu2014,Ozdemir2019}. 
In condensed matter physics, topology entered the stage during the 1980s as of the discovery of the quantum Hall effect~\cite{Klitzing1980}, and is nowadays widely used in materials theory, e.g.,  in terms of topological band theory~\cite{topband}. 
Following the quantum Hall effect, paramount discoveries revealing the importance of topology in condensed matter physics include topological phase transitions and topological matter (which was awarded the 2016 Nobel prize in physics~\cite{nobelprize2016}), such as the gapped topological insulators~\cite{hasankane} and superconductors~\cite{qizhang}, as well as the gapless phases graphene~\cite{goerbig} and Weyl semimetals~\cite{weylreview,HQ2013}. 
The latter is a three-dimensional material, realized in, e.g., TaAs~\cite{XBAN2015,LWFM2015} and TaP~\cite{WSMTaP}, but also in photonic systems~\cite{LWYRFJS2015}, where the valence and conduction band touch at points around which the excitation allow for a theoretical description analogue to that of Weyl fermions. 
These elusive particles were theoretically predicted by Hermann Weyl in 1929 in the context of high-energy physics~\cite{W1929}, and are yet to be observed as fundamental particles. 
In addition to providing a fruitful connection to high-energy physics, this leads to a range of exotic physical phenomena, including the chiral anomaly~\cite{ZXBY2016,NM1983,XPYQ2015,SY2012,G2012,ZB2012,GT2013,PGAPV2014,KGM2015,KGM2017,BKS2018,Huangetal2015} which results in negative magneto-resistance~\cite{AXFT2014,KSXMY2017,Zetal2016}.

The point-like intersections between the valence and conduction bands in Weyl semimetals, naturally dubbed Weyl points or Weyl nodes, are topological; they are characterized by a non-trivial topological invariant, the first Chern number, preventing them from appearing alone in a crystalline material~\cite{topband,weylreview,HQ2013}. 
Thus, every Weyl node characterized by Chern number $+1$ must necessarily be accompanied by a partner characterized by Chern number $-1$. 
Physically, the non-trivial Chern numbers are commonly interpreted as a charge, or chirality, of the corresponding quasi-particle, and the constraint of having a vanishing net charge is referred to as the Nielsen--Ninomiya theorem~\cite{NM1983}. 
This is mathematically equivalent to the Poincar\'e--Hopf theorem, which states that the sum of singularities of a vector field on some manifold is given by the corresponding Euler characteristic~\cite{Poincare,Hopf}. 
This direct mathematical correspondence allows for a description of Weyl semimetals and the Nielsen--Ninomiya theorem in terms of a Mayer--Vietoris sequence of cohomology groups~\cite{Mathai2017a,Mathai2017b}, which further unravels the topological nature of Weyl nodes.

In recent years, successful applications of topology have expanded into the non-Hermitian regime, leading to the emergence of a new research field known as non-Hermitian topological physics~\cite{NHreview}. Despite violating the fundamental axioms of quantum mechanics, non-Hermitian operators find a wide range of applications within both classical and quantum physics; they appear as, e.g., damping matrices~\cite{Fan2022}, reflection matrices~\cite{Topphot}, or effective Hamiltonians~\cite{Lindblad1976}, and are commonly used to describe gain and loss in optical systems~\cite{Ozdemir2019}, friction in mechanical metamaterials~\cite{Ghatak2020}, multi-well dynamics in ultra-cold atoms~\cite{Kreibich2014}, and environment interactions in open quantum systems~\cite{Hatano2019,LiouvEP}.
Relaxing the Hermiticity constraint leads to several unique, yet ubiquitous, physical phenomena, including the breakdown of the conventional bulk-boundary correspondence~\cite{Kunst2018,Yao2018,Edvardsson2018} and the non-Hermitian skin effect~\cite{Ghatak2020}. 
In terms of non-Hermitian topological band theory~\cite{NHTBT}, arguably the most apparent differences from the Hermitian realm are the complex-valued eigenvalues and different sets of left and right eigenvectors of non-Hermitian matrices~\cite{brody14}. 
This leads to the existence of exceptional points (EPs)~\cite{Kato}, eigenvalue degeneracies at which the corresponding eigenvectors coalesce, thereby making the matrix defective and non-diagonalizable.
At EPs, the matrix is rather cast into a Jordan block form, the appearance of which reflects the order of the degeneracy---at an $n$-fold EP  (EP$n$), $n$ eigenvalues and eigenvectors simultaneously coalesce, and the parent matrix host an $n\times n$ Jordan block.
The complex eigenvalues furthermore allow for the possibility for them to braid through the Brillouin zone, sourcing topology exclusively connected to the eigenvalues.
As a consequence, EP$n$s appearing in $m\times m$-matrices, with $m>n$ are non-Abelian in nature, while EP$n$s in $n\times n$-matrices display Abelian topological properties.

EPs are more common than ordinary Hermitian eigenvalue degeneracies. 
Generic EP$n$s are of codimension $2(n-1)$, meaning that their stable appearance requires the simultaneous tuning of $2(n-1)$ parameters~\cite{BerryDeg,NHreview,NHTBT,NHDT,Rotter,Heiss}. 
Thus, stable EP2s appear already in matrices describing two-dimensional systems, in contrast to the Hermitian counterparts that require the simultaneous tuning of three parameters, and thus appear in a stable fashion in three dimensions. 
Three-dimensional non-Hermitian systems instead host potentially linked or knotted contours of EP2s~\cite{carlstroembergholtz,molina,disorderlinesribbons,ourknots,Ronnyknot1,ourknots2,expknots}. 
The codimension of EP$n$s is further reduced for matrices fulfilling some additional similarity relation. 
For example, EP$n$s in pseudo-Hermitian matrices are of codimension $n-1$, while for self skew-similar matrices the codimension depends on the parity of $n$: they are of codimension $n$ for even $n$, and $n-1$ for odd $n$~\cite{Sayyad22,Montag2024}. 
As a consequence, similarity-protected EP2s appear in a stable fashion already in one dimension, while EP3s and EP4s exist in two and three dimensions, respectively~\cite{symprotnod,3EPemil,Delplace21,4EPMarcus,Konig22,Crippa,Sayyad23,Wang2023,Yang2024}.
The ubiquitous existence of EPs has various physical implications, including enhanced sensing~\cite{Hodaei2017,Chen2017,Lai2019,Chu2020,Wiersig2020,Yu2020,Peters2022} and unidirectional lasing~\cite{Peng2014,Feng2014,Hodaei2014,Wang2021}, which highlights their importance.

\begin{figure*}[t!]
\centering

\includegraphics[width=\textwidth]{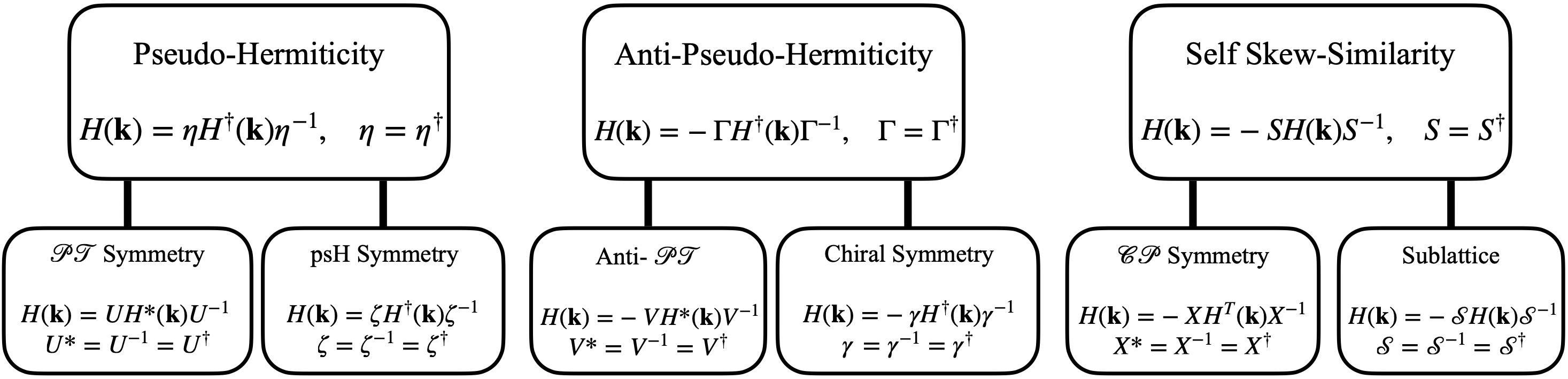}

\caption{\label{fig:Sims} A schematic depiction on the relations between similarity and symmetry relations for non-Hermitian matrices. 
Although symmetries are more directly connected to physical platforms, the topological properties of the eigenvalue degeneracies are more efficiently treated using similarities. 
As the figure indicates, each of the similarities (anti) pseudo-Hermiticity and self skew-similarity covers two physically relevant symmetry relations. 
In terms of codimension of EP$n$s, and their topological classification, the symmetries related to the same parent similarity are deemed equivalent. 
It is however important to note that the corresponding physical implications and interpretations require studies on the specific physical system, for which the symmetry relations are more tractable. 
This picture is inspired by Fig.~1 in Ref.~\cite{Montag2024}, but is modified and reproduced to match current purposes.}
\end{figure*}

Although being abundant in physically relevant systems [recall that pseudo-Hermitian similar matrices include, e.g., parity-time-($\mathcal{PT-}$)symmetric matrices, and self skew-similar matrices include the sublattice-symmetric ones, see Fig.~\ref{fig:Sims}, both widely used in non-Hermitian optics and photonics~\cite{Xiao,Vicencio2015,Mukherjee2015,Topphot}], the topological nature of multifold EPs was just recently uncovered~\cite{Delplace21,Yoshida2024}. 
Similarly to Hermitian Weyl nodes, EP$n$s emerging in $n$-band systems are bound to satisfy an Abelian topological doubling theorem that prevents them from appearing alone in periodic structures.
This topological classification is done by constructing a topological invariant in terms of a {\it resultant winding number}, stemming from the winding of a vector whose components correspond to the resultant of the corresponding characteristic polynomial and its derivatives. 
This comprises the natural extension to EP$n$s in $n$-band systems for arbitrary $n$ of the discriminant number characterizing generic EP2s (by counting the number of times a closed curve around the EP2 crosses the concomitant bulk Fermi arc)~\cite{NHDT}, and the resultant winding numbers characterizing EP3s in Refs.~\cite{Delplace21,Tang2023}.
The resultant winding number is an Abelian topological invariant in the sense that exclusively characterizes the topology associated to EP$n$s appearing in $n\times n$-matrices.

\begin{table*}
\begin{tabular}{|c|c|c|c|c|c|c|c|}
\hline
{\bf Sym.} & {\bf Gen.} &{\bf PsH}& {\bf SSS}& $\mathcal{T}^{\dagger}$/$\mathcal{P}$&$\mathcal{T}$/$\mathcal{I}$&$\mathcal{C}$&$\mathcal{C}^{\dagger}$\\
 \hline
 \multirow{2}{3em}{\bf Codim. EP$n$} & \multirow{2}{3em}{$2n-2$} & \multirow{2}{2em}{$n-1$} & \multirow{2}{2em}{$2\lfloor \frac{n}{2} \rfloor$} & \multirow{2}{3em}{$2n-2$} & \multirow{2}{3em}{$2n-2$} & \multirow{2}{3em}{$2n-2$} &\multirow{2}{3em}{$2n-2$}\\
   & & & & & & &\\
 \hline
\multirow{2}{3em}{{\bf Spec. Const.}}& \multirow{2}{1em}{---} &\multirow{2}{4em}{$\left\{\lambda(\bk)\right\} = \left\{ \lambda^*(\bk)\right\}$}&\multirow{2}{4em}{$\left\{\lambda(\bk)\right\} = \left\{ -\lambda(\bk)\right\}$}&\multirow{2}{4em}{$\left\{\lambda(\bk)\right\} = \left\{\lambda(-\bk)\right\}$}&\multirow{2}{4em}{$\left\{\lambda(\bk)\right\} = \left\{\lambda^*(-\bk)\right\}$}&\multirow{2}{4em}{$\left\{\lambda(\bk)\right\} = \left\{-\lambda(-\bk)\right\}$}&\multirow{2}{4em}{$\left\{\lambda(\bk)\right\} = \left\{-\lambda^*(-\bk)\right\}$}\\
  & & & & & & &\\
  \hline
    \multirow{4}{3em}{{\bf EP$n$ Inv.}}& \multirow{4}{1em}{$\mathbb{Z}$} &\multirow{4}{5em}{$\mathbb{Z}$, $n$ even. $\mathbb{Z}$, $n$ odd. $\mathbb{Z}_2$, $n=2$.} &\multirow{4}{1em}{$\mathbb{Z}$}&\multirow{4}{7em}{$0$, $n=2+4l$. $\mathbb{Z}_2$, $n=3+4l$. $0$, $n=4+4l$. $\mathbb{Z}_2$, $n=5+4l$.}&\multirow{4}{7em}{$\mathbb{Z}$, $n=2+4l$. $0$, $n=3+4l$. $2\mathbb{Z}$, $n=4+4l$. $\mathbb{Z}_2$, $n=5+4l$.}&\multirow{4}{7em}{$0 $, $n=2+4l$. $\text{  }$ $0$, $n=3+4l$. $\text{  }$ $\mathbb{Z}_2$, $n=4+4l$. $\mathbb{Z}_2$, $n=5+4l$.}&\multirow{4}{7em}{$\mathbb{Z}$, $n=2+4l$. $0$, $n=3+4l$. $2\mathbb{Z}$, $n=4+4l$. $\mathbb{Z}_2$, $n=5+4l$.} \\
  & & & & & & &\\
   & & & & & & &\\
   & & & & & & & \\
  \hline
   \multirow{4}{3em}{{\bf Gap Inv.}}& \multirow{4}{1em}{0} &\multirow{4}{1em}{0} &\multirow{4}{1em}{0} &\multirow{4}{7em}{$0$, $n=2+4l$. $\mathbb{Z}_2$, $n=3+4l$. $0$, $n=4+4l$. $\mathbb{Z}_2$, $n=5+4l$.} & \multirow{4}{7em}{$0$, $n=2+4l$. $\text{  }$ $0$, $n=3+4l$.$\text{  }$ $0$, $n=4+4l$. $\mathbb{Z}_2$, $n=5+4l$.} & \multirow{4}{7em}{$0$, $n=2+4l$. $\text{  }$ $0$, $n=3+4l$. $\text{  }$ $\mathbb{Z}_2$, $n=4+4l$. $\mathbb{Z}_2$, $n=5+4l$.} & \multirow{4}{7em}{$0$, $n=2+4l$. $\text{  }$ $0$, $n=3+4l$. $\text{  }$ $0$, $n=4+4l$. $\mathbb{Z}_2$, $n=5+4l$.} \\
  & & & & & & & \\
    & & & & & & &\\
        & & & & & & &\\
  \hline
\end{tabular}
 \caption{A summary of the topological classification of EP$n$s provided in this paper. For each similarity/symmetry, the corresponding spectral constraint is listed, along with the codimension of EP$n$s and the invariants classifying them. Additionally, the topology induced by certain annihilations of EP$n$s is listed, referred to as Gap Inv. Exhibiting non-trivial gap topology indicates that the annihilation of EP$n$s potentially induce bulk Fermi arcs that can only be gapped out by passing through an EP$n$ again; this feature requires the presence of non-local symmetries. Apart from the case of EP2s in systems subject to pseudo-Hermitian similarity (which is classified by a $\mathbb{Z}_2$-invariant), generic EP$n$s and those induced by local similarities are classified by an integer-valued topological invariant---the resultant winding number. More exotic topological features are however induced by EP$n$ emerging at high-symmetry points specified by non-local symmetries; the classification of these depend on the underlying symmetry, but also on the degree of the degeneracy. Above, $\lfloor x \rfloor$ denotes the floor function of $x$.} \label{tab:sumresint}
\end{table*}

\subsection{Summary of results}
In this work, we generalize the topological classification of EP$n$s in $n$-band systems, rewrite it in terms of similarity relations instead of symmetry relations, and extend it to also include EP$n$s protected by self skew-similarity.
Noting that generic EP$n$s, i.e., EP$n$s appearing in the spectrum of matrices without symmetries, are classified by winding numbers in analogy to the topological (Hermitian) symmetry-class AIII, while similarity-protected EP$n$s are classified by either the same winding numbers (EP$n$s protected by self skew-similarity, and odd-fold EP$n$s protected by pseudo-Hermiticity), or Chern numbers (even-fold EP$n$s protected by pseudo-Hermiticity), we are able to reveal a connection to vector bundle classifications previously unknown in non-Hermitian systems.
The classification in terms of Hermitian symmetry classes is possible since the resultant vector around EP$n$s naturally induces a map to a Hermitian Hamiltonian with chiral symmetry [except for EP$n$s with $n$ even protected by (anti) pseudo-Hermitian similarity, where the chiral symmetry is absent].
To further highlight their topological nature, we take advantage of the form of the resultant vector and interpret the doubling theorem for EP$n$s in terms of Mayer--Vietoris sequences within the framework of cohomology theory, which in turn reveal the topological origin of the assigned invariants.
Complementary to this, we also unravel how non-local symmetries enrich the topology of EP$n$s.
This leads to a connection between the non-Hermitian bulk Fermi arcs and Hermitian Dirac strings, suggesting novel topological phases in non-Hermitian $n$-band systems without EP$n$s.
These phases are sourced by (single) topologically protected bulk Fermi arcs induced by non-local symmetries, $\mathbb{Z}_2$-protected Fermi arcs, which are predicted to exist in several $n$-band systems in the absence of EP$n$s.
The main result stemming from the topological classification, including what type of invariants classify the EP$n$-topology and the ``gapped'' topology in the presence of what similarity/symmetry, is summarized in Table~\ref{tab:sumresint}. 
Our results broaden the interpretation of the topological properties of EP$n$s and non-Hermitian band structures in general.
They further make the field of non-Hermitian topological physics accessible to people lacking a background in physics, but rather possess broad knowledge in algebraic topology and geometry.

The rest of the article is structured as follows. In Sec.~\ref{sec:NHDT}, we set the stage by reviewing the non-Hermitian topological doubling theorems derived in Ref.~\cite{Yoshida2024}, rewrite them in terms of similarity relations instead of symmetry relations, and extend it to also include EP$n$s protected by self skew-similarity.
Sec.~\ref{sec:TNEP} further discusses the topological nature of the derived invariants, while Sec.~\ref{sec:VB} connects the topological classification of EP$n$s to vector bundle classification.
Building further on the vector bundle formalism, interpreting the doubling theorems as Mayer--Vietoris sequences is the topic of Sec.~\ref{sec:NHMVS}.Sec.~\ref{sec:NLsym} extends the classification scheme to also include EP$n$s appearing in systems subject to non-local symmetries, including time-reversal ($\mathcal{T}$), particle-hole ($\mathcal{C}$) and inversion ($\mathcal{I}$) symmetries, which further unravels novel topological features relating to how EP$n$s annihilate. Sec.~\ref{sec:COR} discusses corollary results of our general framework.  We conclude, place our results in a wider perspective, and discuss potential future research directions in Sec.~\ref{sec:Con}.

\section{Non-Hermitian Doubling Theorems} \label{sec:NHDT}
This section will be used to set the stage and introduce the concepts of EP$n$s and their concomitant doubling theorems, as well as expressing them in terms of similarities rather than symmetries. 
The doubling theorem explicitly means that EP$n$s of opposite topological invariant emerging within the same $n\times n$ matrix annihilate upon merging.
The classification is further extended to also include EP$n$ protected by self skew-similarity. 
Generic EP$n$s (i.e., EP$n$ appearing in matrices without similarities) are treated in Sec.~\ref{sec:GenEPns}, while similarity-protected EP$n$s are treated in Sec.~\ref{sec:SPEPn}, with Sec.~\ref{sec:psHC} focusing on pseudo-Hermiticity and anti pseudo-Hermiticity, and Sec.~\ref{sec:SSSEPns} on self skew-similarity.

\subsection{Doubling of generic exceptional points} \label{sec:GenEPns}
Consider an $n\times n$ matrix $M(\bk)$, parameterized by $\bk=(k_1,...,k_m)\in \mathbb{T}^m$.
The corresponding eigenvalues are obtained from the characteristic polynomial,
\begin{equation} \label{eq:charpoln}
    P_n(\lambda) = \det{M-\lambda \mathbb{I}} = (-1)^n\left( \lambda^n-\sum_{j=0}^{n-2}a_j\lambda^j\right),
\end{equation}
which is directly written in the retarded form, i.e., setting $\Tr M = 0$~\cite{Sayyad22}. Here, $a_j:\mathbb{T}^m\to \mathbb{C},$ $ \forall j\in\left\{0,...,n-2\right\}$.
From the characteristic polynomial, it can be shown that the coalescence of $l\leq n$ eigenvalues requires solving a set of $2(l-1)$ real equations.
These generically comprise EPs, at which $M$ is not diagonalizable and has a Jordan normal form with a Jordan block of size $l$.
Consequently, the simultaneous tuning of $2(l-1)$ real parameters is required in order for the EPs to appear in a stable fashion. 
In other words, stable $l$-fold EPs are generic in matrices parameterized by $2(l-1)$ parameters, which in this work corresponds to using parameters taking values in $\mathbb{T}^{2(l-1)}$.

A recent work established that EP$n$s appearing in $n\times n$-matrices can be topologically characterized by using a resultant winding number~\cite{Yoshida2024}. 
The argument is based upon the fact that at EP$n$s, the characteristic polynomial has a series of vanishing resultants, namely,
\begin{align} \label{eq:resdef}
r_j&=\text{Res}\left[\partial_{\lambda}^{n-1-j}P_n(\lambda),\partial_{\lambda}^{n-1}P_n(\lambda)\right] = 0, \\
\nonumber
\quad  j&=1,...,n-1.
\end{align}
These resultants will all be directly proportional to the coefficients of the characteristic polynomial,
\begin{equation}
    r_j \propto a_{j-1}, \quad j=1,...,n-1,
\end{equation}
which can be explicitly shown by a straight-forward calculation of the determinant of the Sylvester matrix defining the resultant components in Eq.~\eqref{eq:resdef}.
The different resultants are then collected in a resultant vector as,
\begin{equation}
    \BR = \left[\text{Re}(r_1),\text{Im}(r_1),...,\text{Re}(r_{n-1}),\text{Im}(r_{n-1})\right]^T.
\end{equation}
The topological properties of the EP$n$ are then encoded in a map between $(2n-3)$-spheres. 
Since EP$n$s appear as points in a $2(n-1)$-dimensional parameter space, they can be enclosed by $(2n-3)$-spheres. 
This induces a map from $S^{2n-3}$ in the base space to $S^{2n-3}$ in the space of resultant vectors, i.e., 
\begin{equation}
    \frac{\bk}{|\bk|}\to \frac{\BR}{|\BR|}.
\end{equation}
The degree of this map indicates how many times the map winds around the EP$n$ in the space of resultant vectors, and this winding number can be calculated as
\begin{align}
    W_{2n-3}&=A_{2n-3}\oint_{S^{2n-3}}\Tr \left[\left( \bn ^{-1} d \bn\right)^{(2n-3)}\right], \label{eq:WNRGenRes}
    \\
    A_{2n-3} &= \frac{(n-2)!}{(2\pi i)^{n-1}(2n-3)!},
\end{align}
where $\bn$ is the normalized resultant vector $\bn = \frac{\BR}{|\BR|}$. 
For $n=2$, this coincides with the discriminant winding number~\cite{NHDT}.
Furthermore, this formula makes it apparent that the sum of the winding numbers around all EP$n$s on $\mathbb{T}^{2n-2}$ vanishes. 
To see this explicitly, denote that set of EP$n$s as $\Delta = \{p_1,...,p_k\}$, i.e., there are $k$ distinct EP$n$s on $\mathbb{T}^{2n-2}$.
Using Stokes theorem, the sum of all winding numbers can then be written as
 \begin{widetext}
     \begin{equation}
         \sum_{j=1}^k W^{(k)}_{2n-3} = \sum_{j=1}^k \oint_{S^{2n-3}_k}\Tr \left[\left( \bn ^{-1} d \bn\right)^{(2n-3)}\right] = \int_{\mathbb{T}^{2n-2}\setminus \Delta}d \Tr \left[\left( \bn ^{-1} d \bn\right)^{ (2n-3)}\right] =\int_{\mathbb{T}^{2n-2}\setminus \Delta}d \Tr \left\{\left[d \log\left(\bn\right) \right]^{ (2n-3)}\right\}.
     \end{equation}
 \end{widetext}
The integrand $d \Tr \left[d \log\left(\bn\right) \right]^{(2n-3)}$ vanishes by the second Bianchi identity, which concludes the Abelian doubling theorem for generic EP$n$s in $2(n-1)$ dimensions.

\subsection{Doubling of similarity-protected exceptional points} \label{sec:SPEPn}
When matrices take certain forms, i.e., when they fulfill a similarity relation, the codimension of the EPs of the matrix may decrease. 
Although from a physical point of view, it is more relevant in a direct sense to speak of symmetries instead of similarities, the latter has mathematical benefits as one similarity relation covers several symmetry relations simultaneously~\cite{Montag2024}.
Therefore, the notion of similarities will be employed here.

There are three main classes of similarity relations that affect the codimension of EPs: (Anti) Pseudo-Hermiticity, and self skew-similarity. Fig.~\ref{fig:Sims} shows how these are related to respective symmetries. The similarities are defined as
\begin{widetext}
\begin{alignat}{3}
    \MH_{\text{PH}}(\bk) &= \eta \MH^{\dagger}_{\text{PH}}(\bk)\eta^{-1}, &&\eta=\eta^{\dagger}, \Rightarrow P_{\text{PH}}(\lambda) &&= (-1)^n\left( \lambda^n -\sum_{j=0}^{n-2} a_j\lambda^j\right), \quad a_j\in \mathbb{R}, \label{eq:PHsim}
    \\
    \MH_{\text{aPH}}(\bk) &=-\Gamma\MH_{\text{aPH}}^{\dagger}(\bk) \Gamma^{-1}, \,&&\Gamma=\Gamma^{\dagger}, \Rightarrow P_{\text{aPH}}(\lambda) &&= (-1)^n\left( \lambda^n - \sum_{j=0}^{n-2}a_j\lambda^j\right), \quad \begin{cases} a_j \in i\mathbb{R}, &n+j \text{ odd}, \\ a_j\in \mathbb{R}, &n+j \text{ even},\end{cases}\label{eq:Csim}
    \\
    \MH_{\text{S}}(\bk) &= -S\MH_{\text{S}}(\bk)S^{-1},  &&S=S^{\dagger}, \Rightarrow P_{\text{S}}(\lambda) &&= \begin{cases} (-1)^n\left(\lambda^n-\sum_{j=0}^{n-2} a_{j}\lambda^{j}\right),&a_{2j}\in \mathbb{C}, \,  n \text{ even},\\ 
    (-1)^n\left(\lambda^{n}-\sum_{j=1}^{n-2} a_{2j+1}\lambda^{2j+1}\right),&a_{2j+1}\in \mathbb{C},\, n \text{ odd},  \end{cases} \label{eq:SSSsim}
\end{alignat}
\end{widetext}
with $a_{2j+1}=0$ and $a_{2j}=0$ in Eq.~\eqref{eq:SSSsim} for $n$ even and odd, respectively. These similarities affect the codimension of EPs in different ways. 
Pseudo-Hermiticity [Eq.~\eqref{eq:PHsim}] and anti pseudo-Hermiticity [Eq.~\eqref{eq:Csim}], however, can be considered equivalent from this perspective, and can thus be treated simultaneously. 
For this reason, the term pseudo-Hermiticity will henceforth refer to both anti pseudo-Hermiticity and pseudo-Hermiticity, and we will only specify when necessary. 
Self skew-similarity [Eq.~\eqref{eq:SSSsim}] differs from the two previous ones, mainly because of the appearance of the coefficients in the characteristic polynomial~\cite{Sayyad22}, and will hence be treated separately.

\subsubsection{Pseudo-Hermiticity} \label{sec:psHC}
Under the influence of pseudo-Hermiticity, EP$n$s require only the tuning of $n-1$ real parameters, making them generic in $(n-1)$-dimensional parameterizations of an $n\times n$ matrix. 
These cases are equivalent to the doubling theorems derived for symmetry-protected EP$n$s in Ref.~\cite{Yoshida2024} and are included here for completeness.
For a (anti) pseudo-Hermitian matrix, all components of the resultants will be real (imaginary). 
Hence, the resultant vector components $r_j$ are introduced as
\begin{equation}
    r_j = (-i)\text{Res}\left[\partial^{n-1-j}_{\lambda}P_n(\lambda),\partial^{n-1}_{\lambda}P_n(\lambda)\right],
\end{equation}
where the $(-i)$-factor is only included for relevant components of the anti pseudo-Hermitian matrices. 
This resultant vector then induces a map between $(n-2)$-spheres centered around EP$n$s, the degree of which gives the corresponding winding number associated to the EP$n$. 
The winding number is calculated in a completely analogue way to that for generic EP$n$s,
\begin{equation} \label{eq:WSPEP}
    W_{n-2} \propto \oint_{S^{n-2}} \Tr \left[\left( \bn^{-1} d\bn\right)^{(n-2)}\right].
\end{equation}
The primary difference lies in the dimension of the sphere over which the integral is taken, and a concomitant doubling theorem follows directly.
For $n=3$, this coincides with the resultant winding numbers introduced in Refs.~\cite{Delplace21,NHDT}.
It is however important to point out the case $n=2$ as a special case since the corresponding resultant vector is one-dimensional and hence a scalar. 
Moreover, the integral domain will be $S^0$, which comprises two points. 
Although integration over discrete points is expected to give a trivial result, integration on $S^0$ can be rigorously defined. 
To each point $p$ on $S^0$, one assigns a signature denoted $\sigma(p)$, which evaluates to $\pm 1$. 
The integral over $S^0$ is then defined as the sum of the function values at the respective points, weighted by the corresponding signature. 
In this sense, the above formula is still applicable. 
To illustrate this, it is fruitful to study the example
\begin{equation}
    M_{2\times 2}^{\mathcal{PT}} = \begin{pmatrix}0&1\\k&0\end{pmatrix}, \quad \BR = k\in \mathbb{R}.
\end{equation}
This matrix has an EP2 at $k=0$, around which the winding number is given by,
\begin{equation}
    W = \frac{1}{2} \int_{S^0} \frac{-4k}{4|k|} = -\frac{1}{2} \int_{S^0} \sgn(k).
\end{equation}
Using the above-mentioned definition of integration on $S^0$, one obtains
\begin{equation}
    W = -\frac{1}{2}\left[\sgn(-p)\sigma(-p)+\sgn(p)\sigma(p)\right].
\end{equation} 
Choosing a signature on the points $-p$ and $p$ corresponds to choosing an orientation. 
To stay consistent with the orientation in higher dimensions, $\sigma(\pm p) = \pm 1$, and the winding number becomes $W=-1$. 
Thus, the formula for the winding number Eq.~\eqref{eq:WSPEP} (and all concomitant doubling theorems derived from it)  is applicable for similarity-protected EP$n$s for any $n\geq 2$.
It should be noted, however, that the zero-dimensional winding number does not take integer values, but rather values in $\mathbb{Z}_2$. 
This will be further explained in Sec.~\ref{sec:TNEP}.

\subsubsection{Self skew-similarity} \label{sec:SSSEPns}
In contrast to the doubling theorems discussed above, which can be directly extracted from the work done in Ref.~\cite{Yoshida2024}, EP$n$s appearing in self skew-similar matrices are not covered in that reasoning, which is why they have to be treated separately.
A self skew-similar $n\times n$ matrix has to be studied in a slightly different way since the codimension of the EP$n$s differs depending on the parity of $n$. 
Keeping this in mind, the resultant vector reasoning can be used also for self skew-similar matrices, although it has to be modified.

When $n$ is even, the codimension of an EP$n$ is $n$. Defining the components of the resultant vector $r_j$ as
\begin{align}
    r_{2j-1} = \text{Re}\left\{\text{Res}\left[\partial_{\lambda}^{n-2j}P_n(\lambda),\partial^{n-1}_{\lambda} P_n(\lambda)\right]\right\}, \label{eq:resvecsss1}
    \\
    r_{2j} = \text{Im}\left\{\text{Res}\left[\partial_{\lambda}^{n-2j}P_n(\lambda),\partial^{n-1}_{\lambda} P_n(\lambda)\right]\right\}, \label{eq:resvecsss2}
\end{align}
where $j\in\left\{1,...,n/2 \right\}$, gives a resultant vector $\BR$ with $n$ components, whose singularities correspond exactly to the EP$n$s of the parent self skew-similar matrix. 
As before, the resultant vector induces a map between $(n-1)$-spheres, the degree of which around the EP$n$s defines the respective winding number,
\begin{equation}
    W\propto \oint_{S^{n-1}}\Tr\left[\left(\bn^{-1}d\bn\right)^{(n-1)}\right].
\end{equation}
Their corresponding sum is bound to vanish when occurring on a torus.
When $n$ is odd, the codimension of an EP$n$ is $n-1$, which is the same situation as for pseudo-Hermitian systems. Importantly, the resultant vector has to be introduced in a different manner, due to the different nature of the characteristic polynomial. 
Defining the $(n-1)$ components as in Eqs.~\eqref{eq:resvecsss1} and \eqref{eq:resvecsss2}, gives a resultant vector $\BR$ that induces a map between $(n-2)$-spheres, and the corresponding doubling theorem follows by the same reasoning as in the earlier cases.
This implies that EP$n$s protected by self skew-similarity also obey an Abelian doubling theorem, a result that complements that of Ref.~\cite{Yoshida2024}.

\section{Topological Nature of Resultant Winding Number} \label{sec:TNEP}
Complementary to the previous section, this section deals with the nature of the topological invariant.
Mapping the resultant vector field to a Hermitian Hamiltonian of a topological insulator through the Clifford algebra, establishes an equivalence between the Abelian non-Hermitian eigenvalue topology and the (Hermitian) tenfold way classification of topological matter~\cite{Altland1997, kitaev_periodic_2009, ryu_topological_2010, schnyder_classification_2008}. 
This provides an interpretation of the topological invariants classifying EP$n$s in terms of the tenfold way. 
Since this connection is different for generic and similarity-protected EP$n$s, these will be treated separately in Secs.~\ref{sec:GenEPnTN} and \ref{sec:SPEPnTN}, respectively.

\subsection{Generic EP$n$s} \label{sec:GenEPnTN}
Consider an EP$n$ at $\bm k = 0$. As seen before, this naturally induces a map 
\begin{align}
S^{2n-3} &\to S^{2n-3},\\
    \frac{\bm k}{| \bm k |}&\mapsto \frac{\bm R}{| \bm R |},
\end{align}
which shows how a sphere around the EP$n$ winds around the origin in the resultant space. Up to homotopy, this map is an element of the homotopy group of spheres of the same dimension, which is classified by integers 
\begin{align}
    \pi_{2n-3}\left(S^{2n-3}\right) = \mathbb Z.
\end{align}
The topological invariant can be thought of as the winding number in symmetry class AIII in odd dimensions~\cite{Altland1997,kitaev_periodic_2009, ryu_topological_2010, schnyder_classification_2008}. 
There is a natural mapping to a chirally symmetric resultant Hamiltonian of dimension $2^{n-1}\times 2^{n-1}$,
\begin{equation} \label{eq:resham}
H_{\BR}(\bk) = \sum_{j=1}^{\dim \BR}r_j(\bk)\gamma^{j}.
\end{equation}
Here, $\gamma^i$ are $2n-1$ matrices of dimension $2^{n-1}\times 2^{n-1}$ satisfying $\{\gamma^i , \gamma^j  \} = 2\delta^{ij}$.
Therefore, $H_{\BR}(\bk)$ describes a $2^{n-1}$-band model in the AIII class, where the chiral symmetry operator is given by $\gamma_{2n-1}$.
It should be noted that this is generally not the most generic form of a $2^{n-1}$-band model, but rather it describes a system with a generalized $2^{\lfloor \frac{n}{2} \rfloor -1}$ ``spin''-degeneracy, with $\lfloor x \rfloor$ denoting the integer part of $x$.
The Hamiltonian of such a model can be written in off-diagonal form, and hence the resultant Hamiltonian becomes
\begin{align}
    H_{\BR} = \left(
\begin{array}{cc}
0 & q   \\
q^{\dagger} & 0 
\end{array}
\right).
\end{align}
The higher-dimensional winding numbers are explicitly given by 
\begin{align}
    W_{2n-3} &=A_{2n-3}\oint_{S^{2n-3}}\Tr \left[\left(q^{-1} dq\right)^{(2n-3)}\right],
    \\
    A_{2n-3} &= \frac{(n-2)!}{(2 \pi i)^{n-1}(2n-3)!},
\end{align}
which coincides exactly with Eq.~\eqref{eq:WNRGenRes}. This implies that the resultant winding number associated with a generic EP$n$ can be interpreted as a generalized winding number corresponding to topological invariants of the Hermitian AIII symmetry class.

\subsection{Similarity-protected EP$n$s} \label{sec:SPEPnTN}
\subsubsection{Pseudo-Hermiticity}
For EP$n$s of codimension $n-1$, the topological invariant will have a different meaning depending on the parity of $n$. 
When $n$ is odd, the same reasoning as for generic EP$n$s can be employed, since a Hamiltonian defined as in Eq.~\eqref{eq:resham} will still obey the emergent chiral symmetry. 
The resultant winding number will therefore define a map between $(n-2)$-spheres, which are classified by integers, $\pi_{n-2}(S^{n-2})=\mathbb{Z}$.
This means that the resultant winding number can be considered as topological invariants from symmetry class AIII in odd dimensions. 

When $n$ is even, however, the case will be different. 
The Hamiltonian defined through Eq.~\eqref{eq:resham} will then obey no symmetries since it includes all $\gamma$-matrices in the respective dimension, and will thus be in symmetry class A. 
The resultant vector still defines a map between $(n-2)$-spheres, classified by integers.
For $n>2$, the corresponding topological invariants are therefore Chern numbers instead of winding numbers~\cite{Altland1997,kitaev_periodic_2009, ryu_topological_2010, schnyder_classification_2008}. 
In all generality, this means that EP$n$s will be classified by the $\left(\frac{n}{2}-1\right)$th Chern number for even $n$. 

The case $n=2$ requires additional care since the general reasoning does not fully apply.
The resultant vector still induces a map between spheres, although now between $0$-spheres. 
Since $\pi_0(S^0)=\mathbb{Z}_2$, these maps are not classified by integers, but by some $\mathbb{Z}_2$-valued invariant. 
This can be understood in the following way.
Since $S^0$ consists of two points, maps between different $S^0$ can only be of two different kinds.
Either the two points are sent to different points or to the same, indicating a $\mathbb{Z}_2$-classification.
That one-dimensional systems subject to pseudo-Hermitian similarity are classified by $\mathbb{Z}_2$-invariants is consistent with the $\mathbb{Z}_2$-classification for one-dimensional $\mathcal{PT}$-symmetric systems derived in Ref.~\cite{Yang2024}.
The corresponding resultant Hamiltonian derived from Eq.~\eqref{eq:resham} will be a one-band model without chiral symmetry.
Moreover, restricting it to $S^0=\{p,-p\}$, choosing the signature $\sigma(\pm p)=\pm 1$ as before, the resultant Hamiltonian takes the form
\begin{equation}
    H_{\BR}(k)_{|_{S^0}} = \begin{cases}H_{\BR}(-p)\sigma(-p)&\\H_{\BR}(p)\sigma(p)& \end{cases} = \begin{cases} -H_{\BR}(-p)&\\H_{\BR}(p)&\end{cases}.
\end{equation}
When sending $k\to -k$, the Hamiltonian transforms as follows,
\begin{equation}
    H_{\BR}(-k)_{|_{S^0}} = \begin{cases}H_{\BR}(p)\sigma(-p)&\\H_{\BR}(-p)\sigma(p)& \end{cases} = \begin{cases} -H_{\BR}(p)&\\H_{\BR}(-p)&\end{cases}.
\end{equation}
This implies that $H_{\BR}(k)=-H_{\BR}(-k)$, and consequently, the resultant Hamiltonian satisfies particle-hole symmetry with generator $1$.
This means that the classification falls into symmetry class D, which in the relevant dimension is classified by $\mathbb{Z}_2$-invariants, consistent with all the above reasoning.

\subsubsection{Self skew-similarity}
Since EP$n$s protected by self skew-similarity will be of codimension $n-1$ when $n$ is odd, these are covered by the reasoning above and will give topological invariants belonging to symmetry class AIII, i.e., winding numbers. 
Interestingly, this will also be the case for EP$n$s when $n$ is even. 
This can be seen from the definition of the corresponding resultant vectors.
Due to the appearance of the characteristic polynomial, the resultant vector will have an even number of components both for even and odd values of $n$.
Physically, this can be seen as a consequence of one of the corresponding bands being forced to be flat (i.e., one of the eigenvalues is always 0) when $n$ is odd.
The resultant winding will therefore be given in terms of the degree of a map between spheres of odd dimensions, and maps between $(n-1)$-dimensional spheres classify both EP$n$s and EP$(n+1)$s for even $n$.
This can be interpreted as topological invariants in symmetry class AIII.
This marks a crucial topological difference between similarity-protected EPs of even order; if protected by pseudo-Hermiticity, they are topologically classified by Chern numbers, while those protected by self skew-similarity are topologically classified by winding numbers. 
This furthermore indicates deeper topological phenomena related to the corresponding vector bundle classification, which will be briefly discussed in the following section. 

\begin{figure}[t!]
\centering

\includegraphics[width=\columnwidth]{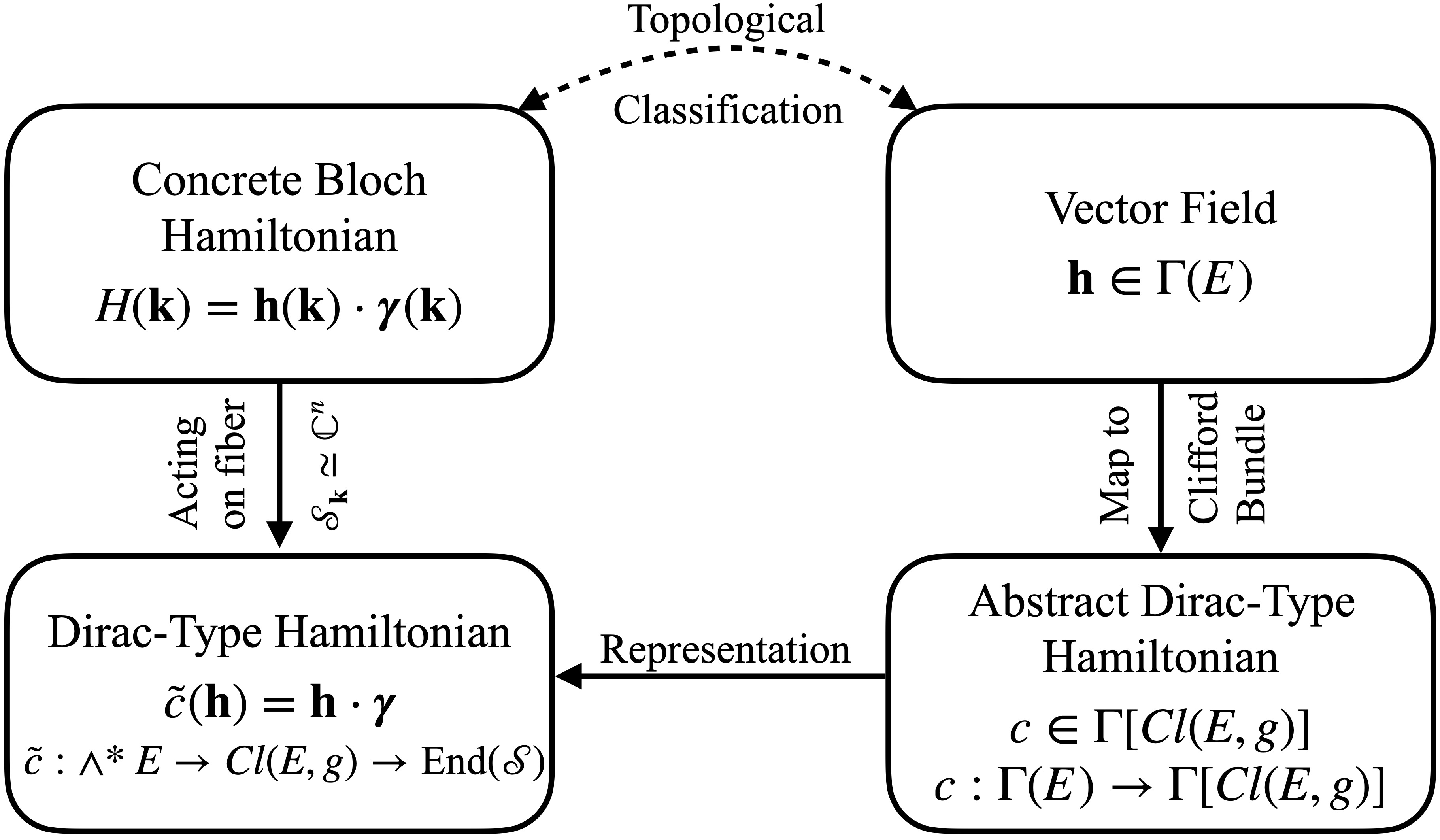}

\caption{A schematic illustration of the importance of the geometric connection between Bloch Hamiltonians and vector bundles. Understanding a Bloch Hamiltonian in terms of its corresponding vector bundle construction allows for a mathematical relation to vector fields. The formal connection to the corresponding vector fields further allows for a more fundamental origin of the topological features widely present in non-Hermitian matrices used to model various systems of physical relevance. The classification scheme will hence comprise a fruitful platform to enrich the theoretical understanding of these systems. For physical systems, $E$ can be chosen as the tangent bundle of the corresponding Brillouin torus, which further enables a choice of constant $\gamma$-matrices. \label{fig:VBC} }
\end{figure}

\section{Vector Bundle Classifications} \label{sec:VB}
This section is devoted to commenting on the more fundamental and abstract picture provided by the above scheme in terms of a more general vector bundle classification.
The key element here is the resultant Hamiltonian in Eq.~\eqref{eq:resham}.
Since the relations between Hamiltonians, vector fields, and vector bundles have been clarified for vector bundles of rank 3 and higher (corresponding to systems in dimension 3 and higher), rank 1 and 2 vector bundles will be treated separately in Secs.~\ref{sec:VBCODIM2} and \ref{sec:VBCODIM1}, respectively. 
Since this will build upon the construction for higher-rank vector bundles, these will be treated first in Sec.~\ref{sec:VBCODIM3}. 
The results of the below reasoning are summarized in Table~\ref{tab:sumres}, and the conducted method is illustrated in Fig.~\ref{fig:VBC}.

\begin{table*}
\begin{tabular}{|c|c|c|c|}
\hline
{\bf EP$n$ Type} & {\bf Generic} &{\bf Pseudo-Hermitian}& {\bf Self Skew-Similar}\\
 \hline
 {\bf Codimension} & $2n-2$ & $n-1$&$2\lfloor \frac{n}{2} \rfloor$\\
 \hline
\multirow{2}{6em}{{\bf Resultant Hamiltonian}}& \multirow{2}{8em}{$H_{\BR} = \sum_{i=1}^{2n-2}r_i \gamma^i$ $\{H_{\BR},\gamma^{2n-1}\}=0$} &\multirow{2}{10em}{$H_{\BR} = \sum_{i=1}^{n-1}r_i \gamma^i$ $\{H_{\BR},\gamma^{n}\}=0$, $n$ odd} &\multirow{2}{9em}{$H_{\BR} = \sum_{i=1}^{2\lfloor \frac{n}{2} \rfloor}r_i \gamma^i$ $\{H_{\BR},\gamma^{2\lfloor \frac{n}{2} \rfloor+1}\}=0$} \\
  & & & \\
  \hline
    \multirow{2}{8em}{{\bf Corresponding Hermitian Deg.}}& \multirow{2}{6em}{$2^{n-1}$-fold protected} &\multirow{2}{12em}{$2^{\frac{n-1}{2}}$-fold, prot., $n$ odd. $2^{\frac{n-2}{2}}$-fold gen., $n$ even.} &\multirow{2}{7em}{$2^{\lfloor\frac{n}{2}\rfloor-1}$-fold protected} \\
  & & & \\
  \hline
   \multirow{3}{7em}{{\bf Topological Invariant}}& \multirow{3}{6em}{Class AIII, $\mathbb{Z}$} &\multirow{3}{10em}{Class A, $n$ even, $\mathbb{Z}$. Class AIII, $n$ odd, $\mathbb{Z}$. Class D, $n=2$, $\mathbb{Z}_2$.} &\multirow{3}{6em}{Class AIII, $\mathbb{Z}$} \\
  & & & \\
    & & & \\
  \hline
  {\bf Vector Bundle}& $T\mathbb{T}^{2n-2}$, rank $2n-2$ &$T\mathbb{T}^{n-1}$, rank $n-1$ &$T\mathbb{T}^{2\lfloor \frac{n}{2} \rfloor}$, rank $2\lfloor \frac{n}{2} \rfloor$ \\
  \hline
  {\bf Vector Field} &$\BR \in \Gamma(T\mathbb{T}^{2n-2})$ &$\BR \in \Gamma(T\mathbb{T}^{n-1})$&$\BR \in \Gamma(T\mathbb{T}^{2\lfloor \frac{n}{2} \rfloor})$ \\
  \hline
\end{tabular}
 \caption{A summary of the topological classification of EP$n$s, including their codimension, relation to Hermitian degeneracies and symmetry classes, and their corresponding vector bundle interpretation. This provides the full topological picture of generic and similarity-protected EP$n$s from the abstract notion of vector bundles to concrete Bloch Hamiltonians via the notion of sections of the vector bundle (or, equivalently, vector fields on the base manifold). The explicit dependence on the parameter $\bk$ has been neglected for brevity, and $\lfloor x \rfloor$ denotes the floor function of $x$.} \label{tab:sumres}
\end{table*}

\subsection{Vector bundle classification of EP$n$s of codimension $\geq 3$} \label{sec:VBCODIM3}
Due to the relation between EP$n$s and Hermitian Hamiltonians expressed in terms of the corresponding resultant vector, the vector bundle classification of EP$n$s with codimension larger than 3 will be analogous to the vector bundle classification of Dirac-like Hamiltonians provided in Ref.~\cite{Mathai2017b}. For completeness, we summarize this classification scheme here, with further details available in Ref.~\cite{Mathai2017b}.

Denote by $E$ an oriented real vector bundle of rank $d$ over a compact oriented base manifold $M$. 
Furthermore, assume that this is accompanied by a spin$^c$ structure and a fiber metric $g$ on $M$.
The spin structure naturally enables the construction of a spinor bundle, usually denoted $\mathcal{S}$, i.e., the Hilbert space. This bundle is sometimes also referred to as the Bloch bundle.
Then there is a map $\tilde{c}$ from the exterior algebra bundle of $E$, $\bigwedge^*E$, via the Clifford algebra bundle $Cl(E,g)$, to the endomorphism group of $\mathcal{S}$, i.e., the group of operators on the Hilbert space (among which are Hamiltonians). 
The map $\tilde{c}$ acts such that it relates sections of $E$, to a Dirac Hamiltonian, i.e.,
\begin{equation} \label{eq:cdo}
    \tilde{c}(\BR) = \BR \cdot \boldsymbol{\gamma}, \quad \BR\in \Gamma(E).
\end{equation}
The Clifford bundle furthermore induces Clifford bracket relations, meaning that an orthonormal frame in $E$, denoted $\{e_i\}_{i=1}^d$, are mapped to $\{c(e_i)\}_{i=1}^d$ in the Clifford bundle. 
These are bound to satisfy $c(e_i)c(e_j)+c(e_j)c(e_i)=2g_{ij}$, which are the usual Clifford bracket relations (this is the conventional way to construct Clifford algebra structures). 
Identifying $\tilde{c}(e_i)=\gamma_i$, the connection to the usual notation becomes more obvious.

Given all these constructions, the line between vector bundles and Dirac Hamiltonians is as follows (the phrasing below is the same as used in Ref.~\cite{Mathai2017b} for the sake of clarity). 
A section of the vector bundle $E$, $\BR\in \Gamma(E)$, induces an abstract Dirac operator, which is a section of the Clifford bundle, $c(\BR) \in \Gamma[Cl(E,g)]$. 
This operator can be {\it represented} by a concrete Dirac operator, defined as in Eq.~\eqref{eq:cdo}. 
This acts on the Bloch bundle. 
Above a point $\bk\in M$, the concrete Bloch Hamiltonian is defined as $H(\bk) = \BR(\bk)\cdot \boldsymbol{\gamma}(\bk)$, which acts on the corresponding fiber of the Bloch bundles, which will be isomorphic to $\mathbb{C}^n$.

For the physical considerations of direct interest in this work, the manifold $M$ is taken to be the Brillouin torus, $\mathbb{T}^d$. 
This simplifies numerous things, and especially it allows for both the Bloch bundle and the vector bundle $E$ to be trivial ($E$ then becomes the tangent bundle of $\mathbb{T}^d$). 
As a consequence the $\gamma$-matrices can be taken constant~\cite{Mathai2017b}.

The framework is now sufficient to extend this classification to EP$n$s of codimension larger than 3. This means that the following EP$n$s are covered here:
\begin{itemize}
    \item Generic EP$n$s for $n\geq 3$,
    \item EP$n$s protected by pseudo-Hermitian similarity for $n\geq 4$, and,
    \item EP$n$s protected by self skew-similarity for $n\geq 4$.
\end{itemize}
The remaining EP$n$s will be treated in subsequent sections.

Let us start with generic EP$n$s. 
The resultant vector is of dimensions $2n-2$, yielding a concrete Bloch Hamiltonian $\BR\cdot\boldsymbol{\gamma}$ over a point $\bk\in \mathbb{T}^{2n-2}$. 
This can be thought of as a parametrization of a concrete Dirac operator, $\BR\cdot\boldsymbol{\gamma}$, with $\BR$ now denoting a vector field, or, formally, a section over the tangent bundle $E$ of $\mathbb{T}^{2n-2}$, completing the connection to vector bundle classifications.

Consider now EP$n$s protected by pseudo-Hermitian similarity. 
Then the same reasoning as above holds, with $2n-2\to n-1$. 
The classification for EP$n$s and EP$(n+1)$s protected by self skew-similarity is recovered by instead letting $2n-2\to n$.

This concludes the vector bundle classification for these EP$n$s, and we now turn to resolving the remaining lower-order cases.

\subsection{Vector bundle classification of EP$n$s of codimension 2} \label{sec:VBCODIM2}
The reasoning for EP$n$s of codimension 2 covers the cases of generic EP2s, all similarity-protected EP3s, and EP2s protected by self skew-similarity (which in some sense are equivalent to generic EP2s).

Naturally, stable EP$n$s of codimension 2 appear in a stable fashion in two-dimensional systems. Therefore the base manifold $M$ is to be taken as two-dimensional in the corresponding vector bundle classification. 
Since the procedure in Ref.~\cite{Mathai2017b} specifically assumed that the vector bundle $E$ is of rank $d\geq 3$, physically motivated by the stability of Weyl nodes in three dimensions, the same procedure must be performed for a two-dimensional $M$. 
For physical reasons, we restrict ourselves to $M=\mathbb{T}^2$ here. 
Consequently, the vector bundle $E$ denotes the tangent bundle of $\mathbb{T}^2$. 
Let $\{e_1,e_2\}$ denote an orthonormal frame for $E$. 
As before, these are sent to the Clifford algebra generators with the map $c$, and for $d=2$, they can be represented by two of the three Pauli matrices acting on the Bloch bundle. 
Sections on $E$, $\BR\in \Gamma(E)$ then define abstract Dirac operators as $c(\BR)\in \Gamma[Cl(E,g)]$, which can be represented as concrete Dirac operators as $\tilde{c}(\BR) = \BR \cdot\boldsymbol{\sigma}$. 
Note that both $\BR$ and $\boldsymbol{\sigma}$ have two components since $E$ is of rank 2. 
Therefore, the Dirac operators constructed in this way are not complete Dirac operators, but they are subject to an emergent symmetry with the symmetry generator represented by the remaining Clifford generator/Pauli matrix [cf. the resultant Hamiltonian in Eq.~\eqref{eq:resham}]. 
The concrete Bloch Hamiltonian above a point $\bk\in \mathbb{T}^2$ can then be written as $H(\bk) = \BR(\bk)\cdot \boldsymbol{\sigma} = r_1(\bk)\sigma^1+r_2(\bk)\sigma^2$, with $\sigma^3$ the symmetry generator. 
This is exactly the form of the resultant Hamiltonian in Eq.~\eqref{eq:resham} when $\text{dim}(\BR) = 2$, meaning that this reasoning establishes the vector bundle picture of the topological classification of EP$n$s of codimension 2.

\subsection{Vector bundle classification of EP$n$s of codimension 1} \label{sec:VBCODIM1}
Lastly, EP2s protected by pseudo-Hermitian similarity require their own treatment since they are of codimension 1. 
In this case, $M=S^1$, and $E$ can again be taken to be the corresponding tangent bundle. 
The resultant Hamiltonian will be a scalar, yielding a one-band model. 
The representation of the Clifford generator, which is to be a self-adjoint endomorphism, will be given by unity $1$. 
The sections of $E$ will be sent to themselves under the map $\tilde{c}$, i.e., $\tilde{c}:\BR \to \tilde{c}(\BR) = \BR\cdot 1 = \BR$, and the concrete Bloch Hamiltonian will coincide with $\BR$.
It should be noted that in contrast to all the other cases, the topological classification of similarity-protected EP2s, and the corresponding vector bundle classification, is not given in terms of some Hermitian band intersection governed by the resultant Hamiltonian, but rather on the zeros of the eigenvalues of a one-band Hamiltonian.
It should be noted that it still corresponds to singularities of a vector field (or, equivalently, sections) on $M$.
This completes the connection between vector bundles and the topological classification of EP$n$s.

\section{Non-Hermitian Mayer--Vietoris Sequences} \label{sec:NHMVS}
The vector bundle classification allows for an alternative and more direct topology-based approach to the arguments presented in Secs.~\ref{sec:NHDT} and \ref{sec:TNEP}. 
The topological nature of the resultant winding number, as well as the doubling theorems for generic and similarity-protected EP$n$s can be understood in a cohomology framework, inspired by earlier works on Weyl semimetals by Mathai and Thiang~\cite{Mathai2017a,Mathai2017b}.
This section merely comprise an alternative description of the topological invariants and the concomitant doubling theorems of EP$n$s, with the advantage that it reflects their formal topological origin.

\begin{figure*}[t!]
\centering

\includegraphics[width=\textwidth]{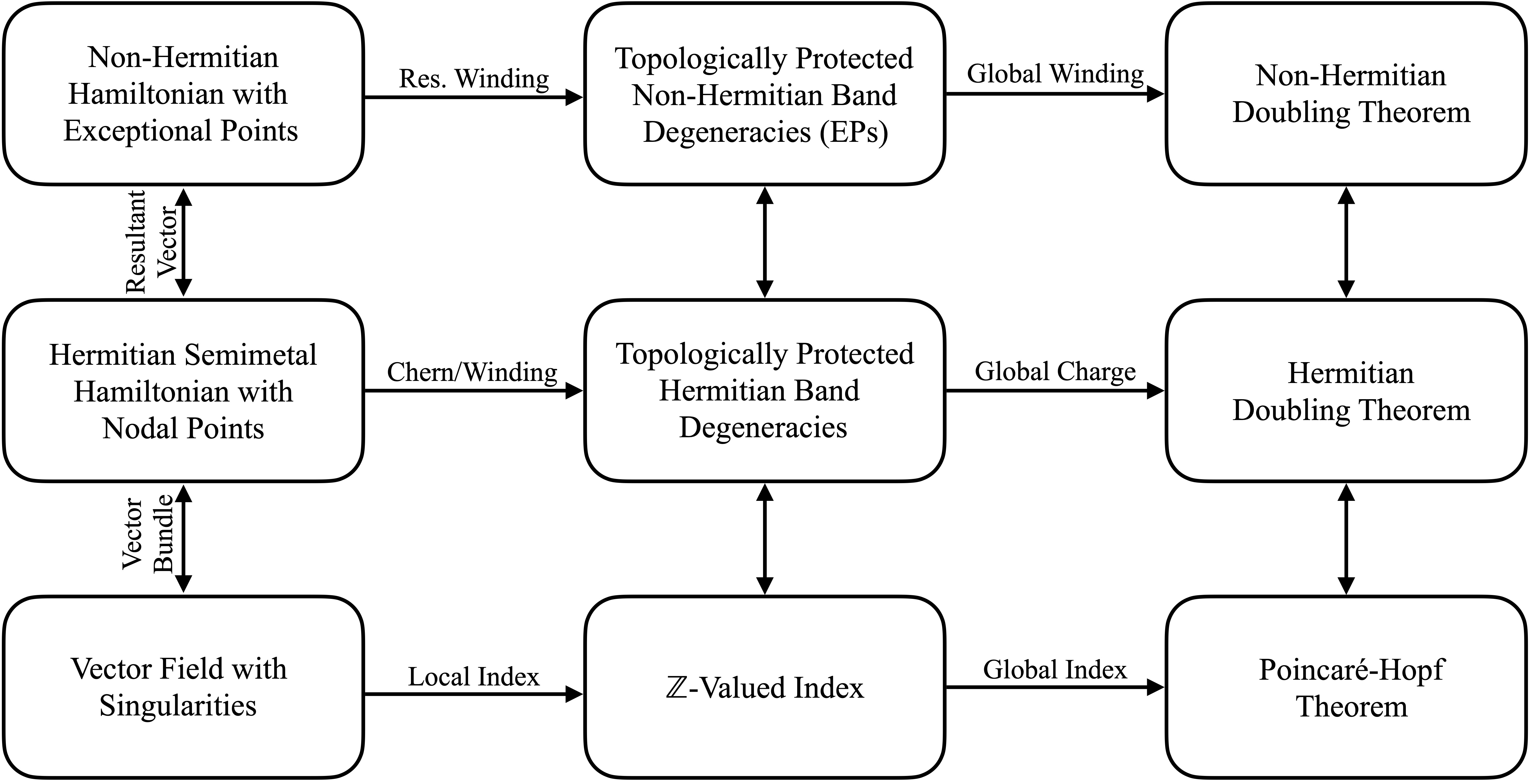}

\caption{A schematic illustration of the relation between the physical and mathematical concepts describing the topological nature of EP$n$s. The key feature is the resultant vector, which is used to translate the non-Hermitian EP$n$ to a Hermitian degeneracy, around which integration domains are well-defined and topological invariants can be consistently calculated. Using (and extending) the results of Mathai and Thiang in Refs.~\cite{Mathai2017a,Mathai2017b}, the connection to vector field singularities becomes apparent. The non-Hermitian doubling theorem can then be understood as a Poincar\'e--Hopf theorem, whose topological origin is reflected in the corresponding Mayer--Vietoris sequences.\label{fig:MPR} }
\end{figure*}

As previously, the different types of EPs will be treated separately for the sake of clarity, starting with generic EP$n$s in Sec.~\ref{sec:GenEPnsMV}, followed by those protected by various similarity relations in Sec.~\ref{sec:SPEPnMV}. The relations between the reasoning in the physical non-Hermitian system (through the non-Hermitian matrices) and the mathematical notions are schematically illustrated in Fig.~\ref{fig:MPR}.

\subsection{Generic EP$n$s} \label{sec:GenEPnsMV}
As a starting point, consider EP2s appearing in $2\times 2$ matrices. This will later be generalized to EP$n$s in $n\times n$ matrices.

  A $2\times 2$ matrix parameterized by $\bk\in \mathbb{T}^2$ has a characteristic polynomial on the form 
\begin{equation}
    P_2(\lambda;\bk) = \lambda^2 - a_0(\bk).
\end{equation}
Suppose $P_2(\lambda;\bk)$ has $l$ eigenvalue degeneracies, which correspond to EP2s.
The resultant vector field whose zeros, or equivalently, singularities, captures the EP2s of the parent matrix, is
\begin{equation}
    \BR(\bk) = \left\{\text{Re}\left[a_0\left(\bk\right)\right],\text{Im}\left[a_0\left(\bk\right)\right]\right\}^T.
\end{equation}
The EP2s can therefore be thought of as isolated singularities of a vector field on $\mathbb{T}^2$, and their topological nature can be described by employing a Mayer--Vietoris argument. By choosing a cover of $\T^2$ as
\begin{equation}
    \mathbb{T}^2=\mathbb{T}^2\setminus\{p_1,...,p_l\} \cup \coprod_{i=1}^l D^2_i,
\end{equation}
meaning that $\T^2$ is punctured $k$ times, corresponding to the singularities, and the punctures are covered with solid disks around the punctures. 
The Mayer--Vietoris sequence in cohomology (with integer coefficients, and denoting $\{p_1,...,p_l\}=:\Delta$) then becomes,
\begin{widetext}
    \begin{equation}
        \dots \to H^d\left(\mathbb{T}^2\right)\to H^d\left(\mathbb{T}^2\setminus \Delta\right)\oplus H^d\left(\coprod_{i=1}^l D_i^2\right) \to H^d\left(\mathbb{T}^2\setminus \Delta \cap \coprod_{i=1}^l D_i^2\right) \to H^{d+1}\left(\mathbb{T}^2\right)\to ... .
    \end{equation}

The part relevant for revealing the topology of EP2s can be extracted by setting $d=1$, giving,
    \begin{equation}
        \dots \to H^1\left(\mathbb{T}^2\right)\to H^1\left(\mathbb{T}^2\setminus \Delta \right)\oplus \underbrace{ H^1\left(\coprod_{i=1}^l D_i^2\right)}_{\simeq 0} \to H^1\left(\mathbb{T}^2\setminus \Delta \cap \coprod_{i=1}^l D_i^2\right) \to H^{2}\left(\mathbb{T}^2\right)\to 0,
    \end{equation}
    \end{widetext}
which put on an explicit form reads,
\begin{equation}
    \dots \to \mathbb{Z}^2 \xrightarrow[]{\alpha}\mathbb{Z}^2\oplus\mathbb{Z}^{l-1} \xrightarrow[]{\beta}\mathbb{Z}^l\xrightarrow[]{\gamma} \mathbb{Z} \xrightarrow[]{\delta}0.
\end{equation}
Let us now try to decipher this. The left-most term describes the topology of one-dimensional slices of the non-punctured torus, which physically means that it is classifying one-dimensional insulator topology. 
Since there are two independent choices of subcircles on $\T^2$ (corresponding to the two different generating cycles of the torus), a topological classification requires a pair of integer invariants. 
Puncturing $\T^2$ yields additional topological properties, with the one-dimensional insulator topology remaining. 
An $l$-punctured $\T^2$ is homotopy equivalent to a wedge sum of $2+l-1$ circles; puncturing $\T^2$ once leaves a space topologically equivalent to a one-dimensional skeleton of $\T^2$, which is a wedge sum of two circles.
Any additional punctures give rise to additional circles in the wedge sum, resulting in an extra factor of $\mathbb{Z}^{l-1}$ in the second term on the left.
The next term defines the local charges, or topological indices, around each puncture, while the rightmost term represents the two-dimensional insulator topology.

Having interpreted all the groups, let us now think about the maps $\alpha, \beta$ and $\gamma$, keeping in mind the sequence is exact [meaning that $\text{ker}(\beta)=\text{im}(\alpha)$, and $\text{ker}(\gamma)=\text{im}(\beta)$].
This means that the one-dimensional insulator topology is still present when puncturing the torus, but it does not affect the individual invariants of the punctures, since $\beta\circ\alpha = 0$.  
Here, $\alpha$ and $\beta$ are given by their respective restrictions on representative differential forms.  
The map $\gamma$ is the sum of each components in $\mathbb{Z}^k$, i.e., it sums all the different topological invariants of the singularities. 
The sequence then tells us that if these singularities exists on a torus, the total sum of their indices must vanish (since $\gamma \circ \beta = 0$). 
Therefore, we can conclude that the sum of the topological indices of the singularities of the vector field defined from the resultants of the characteristic polynomial of the parent non-Hermitian matrix has to vanish.

The above reasoning is directly generalized by considering a characteristic polynomial on the same form as in Eq.~\eqref{eq:charpoln}, which hosts an $n$-fold degeneracy. 
Recall that $n$-fold degeneracies are characterized by a vanishing resultant vector $\bm{R}$, which defines a vector field on $\mathbb{T}^{2n-2}$. 
Its singularities (zeros) correspond exactly to the EP$n$s of the parent $n\times n$ matrix. 
Using a similar covering of $\mathbb{T}^{2n-2}$ as before, i.e., removing points corresponding to the singularities and filling them in with solid $(2n-3)$-balls, the relevant part of the Mayer--Vietoris sequence looks as follows:
\begin{widetext}
    \begin{equation}
        ...\to H^{2n-3}\left(\mathbb{T}^{2n-2}\right) \to H^{2n-3}\left(\mathbb{T}^{2n-2}\setminus \Delta\right)\oplus \underbrace{H^{2n-3}\left(\coprod_{i=1}^lD^{2n-3}_i\right)}_{\simeq 0} \to H^{2n-3}\left(\coprod_{i=1}^l S^{2n-3}_i\right)\to H^{2n-2}\left(\mathbb{T}^{2n-2}\right)\to 0,
    \end{equation}
\end{widetext}
which results in,
\begin{equation}
    ... \to \mathbb{Z}^{2n-2}\xrightarrow[]{\alpha}\mathbb{Z}^{2n-2}\oplus \mathbb{Z}^{l-1} \xrightarrow[]{\beta} \mathbb{Z}^l \xrightarrow[]{\gamma} \mathbb{Z} \to 0.
\end{equation}
The meaning of these different parts is a straightforward generalization of the two-dimensional case, with the most important part being that the reasoning with the topological indices can be directly applied---an index is assigned to every singularity, the sum of which has to vanish globally if the singularities are located on a torus.

\subsection{Similarity-protected EP$n$s} \label{sec:SPEPnMV}
Since the codimension of similarity-protected EP$n$s will be different depending on the similarity, EP$n$ protected by pseudo-Hermiticity will be treated separate from those protected by self skew-similarity.

\subsubsection{Pseudo-Hermiticity}
The initial reasoning below will include similarity-protected EP$n$s for $n\geq 3$. 
The covering of $\T^2$ in terms of a punctured torus and disks of appropriate dimensions does not result in any additional topological properties as long as $H^{d}(D^x)=0$.
However, for $d=0$ it does since $H^0(D^x)\simeq \mathbb{Z}$.
This means that the covering will induce non-physical topological properties when the relevant Mayer--Vietoris sequence includes such terms, which will be the case for similarity-protected EP2s.
Moreover, these are classified by $\mathbb{Z}_2$-invariants, meaning that the cohomology groups with integer coefficients will not be what classifies similarity-protected EP2s.
Thus, the Mayer--Vietoris sequence will not provide any relevant information for similarity-protected EP2s.

Following the same reasoning as for generic EP$n$s, i.e., by covering the Brillouin torus with a punctured torus and disks of appropriate dimension centered around the punctures, the relevant part of the Mayer--Vietoris sequence looks like
\begin{widetext}
    \begin{equation}
        ...\to H^{n-2}\left(\mathbb{T}^{n-1}\right) \to H^{n-2}\left(\mathbb{T}^{n-1}\setminus\Delta\right)\oplus \underbrace{H^{n-2}\left(\coprod_{i=1}^lD^{n-2}_i\right)}_{\simeq 0} \to H^{n-2}\left(\coprod_{i=1}^l S^{n-2}_i\right)\to H^{n-1}\left(\mathbb{T}^{n-1}\right)\to 0.
    \end{equation}
\end{widetext}
Writing this out gives,
\begin{equation}
    ... \to \mathbb{Z}^{n-1}\xrightarrow[]{\alpha}\mathbb{Z}^{n-1}\oplus\mathbb{Z}^{l-1}\xrightarrow[]{\beta}\mathbb{Z}^l\xrightarrow[]{\gamma} \mathbb{Z}\to 0,
\end{equation}
where the maps $\alpha$, $\beta$ and $\gamma$ are defined as previously. 
The exactness of the sequence directly generates a doubling theorem for the topological indices assigned to the singularities, just as for the generic EP$n$s. 

\subsubsection{Self skew-similarity}
Since the topological classification of EP$n$s protected by self skew-similarity for odd $n$ is shown to follow from the topological classification of EP$n$s protected by pseudo-Hermitian similarity, this section will be devoted to the remaining case, namely EP$n$s protected by self skew-similarity for even $n$. 
The Mayer--Vietoris sequence will then take the following form: 
\begin{widetext}
    \begin{equation}
        ...\to H^{n-1}\left(\mathbb{T}^{n}\right) \to H^{n-1}\left(\mathbb{T}^{n}\setminus \Delta\right)\oplus \underbrace{H^{n-1}\left(\coprod_{i=1}^lD^{n-1}_i\right)}_{\simeq 0} \to H^{n-1}\left(\coprod_{i=1}^l S^{n-1}_i\right)\to H^{n}\left(\mathbb{T}^{n}\right)\to 0.
    \end{equation}
\end{widetext}
Writing this out explicitly results in,
\begin{equation}
    ...\to \mathbb{Z}^n \xrightarrow[]{\alpha} \mathbb{Z}^n\oplus \mathbb{Z}^{l-1}\xrightarrow[]{\beta} \mathbb{Z}^l \xrightarrow[]{\gamma} \mathbb{Z} \to 0,
\end{equation}
where the maps $\alpha$, $\beta$ and $\gamma$ are defined as previously. Again, the exactness of the sequence yields a doubling theorem for the topological indices assigned to the singularities. 

This implies that the doubling theorems for generic and similarity-protected EP$n$s have a direct interpretation in terms of the exactness of Mayer--Vietoris sequences, which further enriches their topological interpretation.

\section{Non-Local Symmetries: Extension of Classification Scheme} \label{sec:NLsym}
Apart from the similarities acting locally in momentum space, there are several other similarities and symmetries that acts non-locally in momentum space. 
These are commonly referred to as time-reversal symmetry ($\mathcal{T}$), particle-hole symmetry ($\mathcal{C}$), parity ($\mathcal{P}$) symmetry and inversion ($\mathcal{I}$) symmetry~\cite{Sayyad22}. 
These symmetries do not affect the codimension of EP$n$s, but their presence have a significant impact on the topological features of systems as they give rise to high-symmetry points. 
EP$n$s emerging at these points do not fall into the same classification scheme as those emerging away from the high-symmetry points.
The subject of this section will therefore be to clarify how the topology of such EP$n$s is affected under the influence of these symmetries. 
For the sake of clarity, the treatment will be separated such that the symmetries that affect the eigenvalue topology differently are treated separately.

The results of this section will be derived using the following representation of the $\gamma$-matrices:
\begin{align}
    \gamma^1 &= \sigma_1\otimes\mathbb{I}\otimes\mathbb{I}\otimes\dots \nonumber
    \\
    \gamma^2 &= \sigma_2\otimes\mathbb{I}\otimes\mathbb{I}\otimes\dots \nonumber
    \\
    \gamma^3&= \sigma_3\otimes \sigma_1\otimes\mathbb{I}\otimes\dots
    \\
    \gamma^4 &= \sigma_3\otimes \sigma_2\otimes\mathbb{I}\otimes\dots \nonumber
    \\
    \gamma^5 &= \sigma_3\otimes \sigma_3\otimes\sigma_1\otimes\dots \nonumber
    \\
    \vdots& \nonumber
\end{align}
In terms of these, the charge-conjugation matrices take the form
\begin{align}\label{eq:n3symmetries}
    C_+ &= \sigma_1\otimes\sigma_2\otimes\sigma_1\otimes\dots,\\
     C_- &= \sigma_2\otimes\sigma_1\otimes\sigma_2\otimes\dots,
\end{align}
which satisfies
\begin{equation}
    \left(\gamma^{n}\right)^ *  = \pm C_\pm \gamma^{n} C_\pm^{-1}= \begin{cases} \gamma^n, &\quad n \text{ odd}, \\ -\gamma^{n}, &\quad n \text{ even}, \end{cases}
    \end{equation}
meaning that $C_+$ flips the sign of even-numbered $\gamma$-matrices, while $C_-$ flips the sign of the odd-number $\gamma$-matrices.

\subsection{The impact of high-symmetry points}
Before starting to investigate the impacts of non-local symmetries specifically, we need to emphasize the importance and impact high-symmetry points [also referred to as time-reversal-symmetric-momenta (TRIM) in the context of $\mathcal{T}$ symmetry].
How these points affect classification schemes of band degeneracies is well-studied within the Hermitian regime~\cite{hasankane}, and the same principles are also valid in non-Hermitian systems.
Generally, a system subject to some symmetry allows for the existence of EP$n$s anywhere in the Brillouin zone, both at and away from the high-symmetry points.
The topology of EP$n$s emerging away from high-symmetry points is still classified in terms of the resultant winding number, and these are still constrained to obey a doubling theorem.
The reason for this is that there is no way to enclose such an EP$n$ (and such an EP$n$ alone) by a surface on which the symmetry is preserved.
The topological invariant of such an EP$n$ is hence calculated on a symmetry-breaking surface, and can therefore not be considered to be intrinsically related to the presence of the symmetry.

If an EP$n$ instead emerge exactly at a high-symmetry point, it can generally be enclosed by a surface on which the symmetry is preserved.
Such EP$n$s might hence differ topologically from those emerging elsewhere in the Brillouin zone.
Again, it is fruitful to make the comparison to $\mathcal{T}$-symmetric Weyl semimetals, in which Weyl nodes away from the TRIM points obey the conventional Nielsen-Ninomiyah theorem, but give rise to an additional $\mathbb{Z}_2$-topology when emerging at the TRIM points.

When talking about the impact on the topological properties of the EP$n$s by the symmetries, we are exclusively referring to the topology of EP$n$s emerging at the high-symmetry points, since the topology of the remaining EP$n$s are already contained within the classification scheme developed previously.

\subsection{Time-reversal$^{\dagger}$ and Parity symmetry} \label{sec:TRSD}

In non-Hermitian systems, there exist two different $\mathcal{T}$ symmetries since taking complex conjugation and matrix transposition of a non-Hermitian matrix are not equivalent actions. This gives rise to a $\mathcal{T}^{\dagger}$ symmetry, which affects the characteristic polynomial in the same way as $\mathcal{P}$ symmetry does. These symmetries are defined as
\begin{align}
    \mathcal{C}_+H^T(\bk)\mathcal{C}_+^{-1} &= H(-\bk),
    \\
    \mathcal{P} H(\bk) \mathcal{P}^{-1} &=H(-\bk),
\end{align}
and their corresponding characteristic polynomials will look like,
\begin{align}
    P_{\mathcal{C}_+/\mathcal{P}}(\lambda;\bk) &= (-1)^n\left(\lambda^n - \sum_{j=0}^{n-2}a_j(\bk)\lambda^j\right),\nonumber
    \\
    a_j(\bk)&\in \mathbb{C}, \quad a_j(\bk) = a_j(-\bk). \label{eq:charpoltrsd}
\end{align}

Since the parent non-Hermitian symmetry is non-local in momentum space, it will be reflected as an emergent symmetry in the resultant vector, which reads,
\begin{equation}
    r_{2m+1}(\bk) = \text{Re}\left[a_{m}(\bk)\right], \quad r_{2m+2}(\bk) = \text{Im}\left[a_m(\bk)\right].
\end{equation}
The parent non-Hermitian $\mathcal{T}^{\dagger}$ and $\mathcal{P}$ symmetries give rise to an emergent symmetry in the resultant vector,
\begin{equation} \label{eq:resvectrsd}
    r_j(\bk) = r_j(-\bk),
\end{equation}
i.e., the resultant vector components are necessarily even functions of momentum if its parent non-Hermitian system is $\mathcal{T}^{\dagger}$ or $\mathcal{P}$-symmetric.
Before introducing the resultant Hamiltonian, it is fruitful to take a second look at Eqs.~\eqref{eq:charpoltrsd} and \eqref{eq:resvectrsd}.
Assuming the existence of and EP$n$ at some $\bk = \bk_1$, the symmetry enforces the existence of a ``symmetric partner'' at $\bk=-\bk_1$.
Around these, the characteristic polynomials and the resultant vectors are equivalent.
As a consequence, these two EP$n$s necessarily have the same resultant winding number; $\mathcal{T}^{\dagger}$ and $\mathcal{P}$-symmetric partners of EP$n$s are equivalent.
By the doubling theorem, stating that the sum of all winding numbers must vanish when taken over the full Brillouin zone, there must be an additional $\mathcal{T}^{\dagger}$ or $\mathcal{P}$-symmetric pair of EP$n$ with opposite winding numbers; the minimal number of EP$n$ is increased from 2 to 4 as a consequence of the symmetry.
Similar features are well-established for Weyl semimetals subject to $\mathcal{T}$ symmetry, where the minimal number of topologically non-trivial Weyl nodes is known to be 4 when $\mathcal{T}$ squares to $-1$~\cite{weylreview}.

Employing the resultant Hamiltonian introduced in Eq.~\eqref{eq:resham}, the eigenvalue topology is governed by
\begin{equation}
    H_{\BR}(\bk) = \sum_{j=1}^{2n-2}r_j(\bk)\gamma^j.
\end{equation}
The symmetry in the resultant vector corresponds to a Hamiltonian with both $\mathcal{T}$ and $\mathcal{C}$ symmetry, given by the charge-conjugation matrices $C_+$ and $C_-$, respectively, since,
\begin{align}
    C_{\pm} H^*_{\BR}(\bk) C_{\pm}^{-1} &=   C_{\pm}\sum_{ j=1}^{2n-2}\left[ r_{j}(\bk) \left(\gamma^j\right)^*\right]C^{-1}_{\pm} \nonumber
    \\
    &=\pm\sum_{ j =1}^{2n-2}\left[ r_{j}(\bk) \gamma^j\right]= \pm\sum_{ j =1}^{2n-2}\left[ r_{j}(-\bk) \gamma^j\right]\nonumber
    \\
    &=\pm H_{\BR}(-\bk).
\end{align}
The corresponding symmetry class to which this resultant Hamiltonian belongs depends on the EP$n$ in the parent non-Hermitian model. 
A summary of this is found in Table~\ref{tab:nonloc}. 

Due to the properties of the charge conjugation matrices for different values of $d=2n-2$, different additional topological features will emerge depending on the value of $n$. 
Two and four band models fall into class CI and DIII, respectively, and are hence expected to not yield any additional topology, whereas three and five band models belong to class CII and BDI, respectively, meaning that they allow for a $\mathbb{Z}_2$-invariant.
For larger $n$, Bott periodicity provides the complete picture for arbitrary $n$: the symmetry classes are periodic in $d$ with a period of $8$, meaning that the classification of $n$ band models is periodic in $n$ with a period of $4$. 

\subsection{Particle-hole symmetry}

In non-Hermitian systems, particle-hole ($\mathcal{C}$) symmetry is defined as
\begin{equation}
    \mathcal{C}_-H^T(\bk)\mathcal{C}_-^{-1} = -H(-\bk),
\end{equation}
and affects the characteristic polynomial as
\begin{align}
    P_{\mathcal{C}}(\lambda;\bk) &= (-1)^n\left(\lambda^n - \sum_{j=0}^{n-2}a_j(\bk)\lambda^j\right)\nonumber
    \\
    a_j(\bk)&\in \mathbb{C}, \quad a_j(\bk) = (-1)^{n+j}a_j(-\bk).
\end{align}
The constraints on the coefficients in the characteristic polynomial will result in different cases depending on the parity of $n$.
This will have a significant impact on the topology, and we will therefore treat even and odd $n$ separately.
\subsubsection{The case of an odd number of bands}
When $n$ is odd, the coefficients of the characteristic polynomial are bound to satisfy the following relations:
\begin{align}
    \text{Re}\left[a_j(\bk)\right] &= \begin{cases} -\text{Re}\left[a_j(-\bk)\right], &j \text{ even}, \\ \text{Re}\left[a_j(-\bk)\right], &j \text{ odd},\end{cases}
    \\
    \text{Im}\left[a_j(\bk)\right] &= \begin{cases} -\text{Im}\left[a_j(-\bk)\right], &j \text{ even}, \\ \text{Im}\left[a_j(-\bk)\right],&j \text{ odd},\end{cases}
\end{align}
meaning that the real and imaginary part are both either even or odd in $\bk$.
Defining the components of the resultant vector as
\begin{equation}
r_{m+1}(\bk) = \text{Re}\left[a_m(\bk)\right], \quad r_{m+n-1}(\bk) = \text{Im}\left[a_m(\bk)\right],
\end{equation}
i.e., by placing all the real parts after each other, followed by the imaginary parts.
The parent non-Hermitian $\mathcal{C}$ symmetry is reflected in the resultant vector as
\begin{equation}
    r_{2m}(\bk) = r_{2m}(-\bk), \quad r_{2m-1}(\bk) = -r_{2m-1}(-\bk).
\end{equation}
The corresponding resultant Hamiltonian,
\begin{equation}
    H_{\BR}(\bk) = \sum_{j=1}^{2n-2}r_{j}(\bk) \gamma^{j},
\end{equation}
will therefore satisfy,
\begin{align}
    H_{\BR}^*(-\bk) &= \sum_{j=1}^{2n-2}r_j(-\bk)\left(\gamma ^j\right)^* \nonumber
    \\
    &= \sum_{j=1}^{2n-2}(-1)^{j}r_{j}(\bk) (-1)^{j+1}\gamma^j \nonumber
    \\
    &= \begin{cases} -H_{\BR}(\bk),
    \\
    C_+C_- H_{\BR}(\bk) \left(C_+C_-\right)^{-1}. \end{cases}
\end{align}
Thus, the resultant Hamiltonian is $\mathcal{C}$-symmetric with unity as generator, and $\mathcal{T}$-symmetric with generator $C_+C_-$, leaving it in symmetry class BDI.

\subsubsection{The case of an even number of bands}
When $n$ is even, the coefficients of the characteristic polynomial instead satisfy:
\begin{align}
    \text{Re}\left[a_j(\bk)\right] &= \begin{cases} \text{Re}\left[a_j(-\bk)\right], &j \text{ even}, \\ -\text{Re}\left[a_j(-\bk)\right], &j \text{ odd},\end{cases} \label{eq:charpolPHS1}
    \\
    \text{Im}\left[a_j(\bk)\right] &= \begin{cases} \text{Im}\left[a_j(-\bk)\right], &j \text{ even}, \\ -\text{Im}\left[a_j(-\bk)\right],&j \text{ odd}.\end{cases} \label{eq:charpolPHS2}
\end{align}
For even $n$, there will always be a larger number components of the resultant vector even in $\bk$.
Hence, the previous representation of the charge conjugation matrices cannot be used to identify the emergent symmetries of the resultant Hamiltonian.
For this particular case the following representation will be used instead:
\begin{equation}
\hat{C}_+ = \left(\sigma_1\right)^{\otimes 2n-2}, \quad \hat{C}_- = \left(\sigma_2\right)^{\otimes 2n-2}.
\end{equation}
This means that $\hat{C}_+$ swaps the sign on $\gamma$-matrices with index $j\in \{2,3,6,7,10,11,14,15...\}$, while $\hat{C}_-$ swaps the sign on $\gamma$-matrices with index $j\in \{ 1,4,5,8,9,12,13,...\}$.
With this representation of the charge-conjugation matrices, the resultant vector is defined as,
\begin{equation}
r_{2m+1}(\bk) = \text{Re}\left[a_m(\bk)\right], \quad r_{2m+2}(\bk) = \text{Im}\left[a_m(\bk)\right].
\end{equation}
Recalling that conjugation swaps the sign of even-indexed $\gamma$-matrices, $H^*_{\BR}(-\bk)$ will take the form
\begin{equation}
    H^*_{\BR}(-\bk) = \sum_{j\in I_-}r_j(\bk)\gamma^j - \sum_{j\in I_+}r_j(\bk) \gamma^j,
\end{equation}
meaning that the sign is swapped on the terms with index $j\in I_- = \{2,3,6,7,10,11,14,15,...\}$, while it is preserved for $j\in I_+ = \{1,4,5,8,9,12,13,...\}$.
Recalling how $\hat{C}_{\pm}$ acts on the $\gamma$-matrices, we can conclude,
\begin{equation}
    \hat{C}_{\pm}H^*_{\BR}(-\bk) \left(\hat{C}_{\pm}\right)^{-1} = \pm H_{\BR}(\bk),
\end{equation}
so the resultant Hamiltonian is $\mathcal{T}$-($\mathcal{C}$-) symmetric with $C_+$ ($C_-$) as generator, leaving it in symmetry class CI.

To recap, the eigenvalue topology of non-Hermitian systems subject to $\mathcal{C}$ symmetry is classified within symmetry class CI when the number of bands is even, and in symmetry class BDI when the number of bands is odd, as displayed in Table~\ref{tab:nonloc}.
Hence, the classification scheme predicts symmetry-induced topology when the number of bands is $n=4,5 + 4m$, for some positive integer $m$.
Below, the cases $n=3,4,5$ will be discussed separately.
We emphasize that the case of $n=2$ is equivalent to that of $\mathcal{T}^{\dagger}$ symmetry treated in Sec.~\ref{sec:TRSD}.

\subsection{The remaining symmetries}
Since the remaining symmetries eventually result in the same topological classification, they are treated within the same subsection.
We will, however, explicitly show this claim by deriving the resultant Hamiltonians corresponding to each symmetry separately. 

\subsubsection{Time-reversal and Inversion symmetry}
In terms of eigenvalue topology, the influence of $\mathcal{T}$ and $\mathcal{I}$ symmetry will be similar because of how they affect the corresponding characteristic polynomial. The different symmetries are defined as
\begin{align}
    \mathcal{T}_+H^*(\bk) \mathcal{T}_+^{-1} &= H(-\bk),
    \\
    \mathcal{I}_-H^{\dagger}(\bk)\mathcal{I}_-^{-1} &= H(-\bk).
\end{align}
The corresponding characteristic polynomials for systems subject to any of these symmetries will be on the form
\begin{align}
    P_{\mathcal{T}_+/\mathcal{I}_-}(\lambda;\bk) &= (-1)^n\left(\lambda^n - \sum_{j=0}^{n-2}a_j(\bk)\lambda^j\right)\nonumber
    \\
     a_j(\bk)&\in \mathbb{C}, \quad a_j(\bk) = a^*_j(-\bk).
\end{align}
Since $a_j(\bk)=a^*_j(-\bk)$, it follows that the real parts of $a_j(\bk)$ are even functions of $\bk$, while the imaginary parts are odd functions of $\bk$. 
The resultant vector components,
\begin{align}
r_{2m+1}(\bk) &= \text{Re}\left[a_m(\bk)\right], \\
r_{2m+2}(\bk) &= \text{Im}\left[a_m(\bk)\right],
\end{align}
are bound to obey
\begin{align} \label{eq:resvecTI1}
    r_{2m+1}(\bk) &= r_{2m+1}(-\bk), \\
    r_{2m+2}(\bk) &= -r_{2m+2}(-\bk). \label{eq:resvecTI2}
\end{align}
The corresponding resultant Hamiltonian reads
\begin{equation}
    H_{\BR}(\bk) = \sum_{j=1}^{2n-2} r_j(\bk) \gamma^j, 
\end{equation}
with $r_j(\bk)$ satisfying Eqs.~\eqref{eq:resvecTI1} and \eqref{eq:resvecTI2}. 
This means that $H_{\BR}(\bk)$ obeys both Hermitian $\mathcal{T}$ and $\mathcal{C}$ symmetry with $\mathbb{I}$ and $C_+C_-$ as generators, respectively, i.e.,
\begin{align}
    H_{\BR}(\bk) &=H^*_{\BR}(-\bk),
    \\
    H_{\BR}(\bk) &=-C_+C_-H_{\BR}^*(-\bk)\left(C_+C_-\right)^{-1}.
\end{align}
The topological classification of these systems is summarized in Table~\ref{tab:nonloc}. 
An important difference compared to the previous case is that the resultant Hamiltonian will always be in symmetry class BDI---the behavior of the symmetry generators does not change preserving the symmetry class. 
Consequently, the topological invariant will instead differ vastly, taking $\mathbb{Z}$-values for EP2s, even numbers for EP4s, $\mathbb{Z}_2$-values for EP5s, while EP3s are predicted to be topologically trivial.

\begin{table*}[hbt!]
\centering
\begin{tabular}{|cc|ccccc|ccccc|ccccc|}
\hline
\multicolumn{2}{|c|}{} 
 & \multicolumn{5}{c|}{$\mathcal{T}^{\dagger},\mathcal{P}$} 
 & \multicolumn{5}{c|}{$\mathcal{T},\mathcal{I}, \mathcal{C}^\dagger$} 
 & \multicolumn{5}{c|}{$\mathcal{C}$} 
\\\hline
$n$ & $d$ 
 & $\mathcal{TT}^*$ & $\mathcal{CC}^*$ & Class & $\pi_{d-1}$ & $\pi_d$ 
 & $\mathcal{TT}^*$ & $\mathcal{CC}^*$ & Class & $\pi_{d-1}$ & $\pi_d$
 & $\mathcal{TT}^*$ & $\mathcal{CC}^*$ & Class & $\pi_{d-1}$ & $\pi_d$
\\\hline
2 & 2 & $+1$ & $-1$ & CI  & $0$     & $0$ 
       & $+1$ & $-1$ & BDI  & $\mathbb{Z}$           & $0$
       & $+1$ & $-1$ & CI  & $0$           & $0$ 
\\\hline
3 & 4 & $-1$ & $-1$ & CII  & $\mathbb{Z}_2$             & $\mathbb{Z}_2$ 
       & $-1$ & $-1$ & BDI & $0$ & $0$
       & $+1$ & $+1$ & BDI & $0$           & $0$
\\\hline
4 & 6 & $-1$ & $+1$ & DIII  & $0$   & $0$ 
       & $-1$ & $+1$ & BDI& $2\mathbb{Z}$           & $0$
       & $+1$ & $-1$ & CI  & $\mathbb{Z}_2$ & $\mathbb{Z}_2$
\\\hline
5 & 8 & $+1$ & $+1$ & BDI  & $\mathbb{Z}_2$ & $\mathbb{Z}_2$ 
       & $+1$ & $+1$ & BDI & $\mathbb{Z}_2$ & $\mathbb{Z}_2$
       & $+1$ & $+1$ & BDI & $\mathbb{Z}_2$ & $\mathbb{Z}_2$
\\\hline
$\vdots$ & $\vdots$ & \multicolumn{5}{c|}{$\vdots$} 
              & \multicolumn{5}{c|}{$\vdots$} 
              & \multicolumn{5}{c|}{$\vdots$}
\\\hline
\end{tabular}
\caption{Symmetry classes classifying the eigenvalue topology of $n$-band systems subject to non-local symmetries. 
The classification yields three different scenarios: one common for $\mathcal{T}^{\dagger}$ and $\mathcal{P}$ symmetry, one common for $\mathcal{T}$, $\mathcal{I}$ and $\mathcal{C}$ symmetry, and one for $\mathcal{C}^{\dagger}$. 
$\pi_{d-1}$ describes symmetry-induced eigenvalue topology from EP$n$s appearing at the high symmetry-points in terms of $(d-1)$-dimensional cuts not including these EP$n$s (very much in analogy to how Chern numbers for Weyl nodes are defined~\cite{weylreview}).
$\pi_d$ describes symmetry-induced topology of the full $d$-dimensional system when the EP$n$s are gapped out, and therefore describes the possibility of having topologically non-trivial bulk Fermi arcs, $\mathbb{Z}_2$-protected Fermi arcs, in the spectrum, something that is explained in Sec.~\ref{sec:NLPhys}.
These require moving through a phase including an EP$n$ in order to be removed from the spectrum, while trivial bulk Fermi arcs can be ``gapped'' out without passing through such a phase.
For $\mathcal{T}^{\dagger}$ and $\mathcal{P}$ symmetries, these exists when the number of bands are odd, while $\mathcal{C}$ symmetry requires $4+4m$ and $5+5m$ number of bands for their existence.
The remaining symmetries, $\mathcal{T}$, $\mathcal{I}$ and 
$\mathcal{C}^{\dagger}$ symmetry, this feature is unique to $5+4m$ number of bands.
}
\label{tab:nonloc}
\end{table*}

\subsubsection{Particle-hole$^{\dagger}$ symmetry}
Just as for $\mathcal{T}$ symmetry, there exist two different kinds of $\mathcal{C}$ symmetries for non-Hermitian systems. In addition to the ordinary $\mathcal{C}$, there is also $\mathcal{C}^{\dagger}$ defined as 
\begin{equation}
    \mathcal{T}_-H^*(\bk)\mathcal{T}_-^{-1} = -H(-\bk),
\end{equation}
with a characteristic polynomial on the form
\begin{align}
    P_{\mathcal{C}^{\dagger}}(\lambda;\bk) &= (-1)^n\left(\lambda^n - \sum_{j=0}^{n-2}a_j(\bk)\lambda^j\right)\nonumber
    \\
     a_j(\bk)&\in \mathbb{C}, \quad a_j(\bk) = (-1)^{n+j}a_j^*(-\bk).
\end{align}
Just as for ordinary $\mathcal{C}$, the coefficients of the characteristic polynomial will behave differently depending on the parity of $j$ and $n$. Starting again by assuming $n$ to be even, $a_j(\bk)$ obey
\begin{align}
    \text{Re}\left[a_j(\bk)\right] &= \begin{cases} \text{Re}\left[a_j(-\bk)\right], &j \text{ even}, \\ -\text{Re}\left[a_j(-\bk)\right], &j \text{ odd},\end{cases}
    \\
    \text{Im}\left[a_j(\bk)\right] &= \begin{cases} -\text{Im}\left[a_j(-\bk)\right], &j \text{ even}, \\ \text{Im}\left[a_j(-\bk)\right],&j \text{ odd}.\end{cases}
\end{align}
By defining the resultant vector as
\begin{align}
r_{m+1}(\bk) &= \text{Re}\left[a_m(\bk)\right], \\ 
r_{n-1+m}(\bk) &= \text{Im}\left[a_{n-2-m}(\bk)\right],
\end{align}
the $\mathcal{C}^{\dagger}$ symmetry is reflected in the resultant vector components as,
\begin{align}
r_{2m+1}(\bk) &= r_{2m+1}(-\bk),
\\ 
r_{2m+2}(\bk) &= -r_{2m+2}(-\bk),
\end{align}
i.e., the odd numbered components are even functions of momentum, while the even components are odd functions of momentum.
Therefore, the following holds for the resultant Hamiltonian:
\begin{align}
H^*_{\BR}(-\bk) &= \sum_{j=1}^{2n-2} r_j(-\bk)\left(\gamma^j\right)^* =\sum_{j=1}^{2n-2} r_j(\bk) \gamma^j \nonumber
\\
&=\begin{cases} H(\bk), \\ -C_+C_-H(\bk)\left(C_+C_-\right)^{-1}, \end{cases}
\end{align}
which means that $H_{\BR}(\bk)$ is $\mathcal{T}$- and $\mathcal{C}$-symmetric with generators $\mathbb{I}$ and $C_+C_-$, respectively.

When $n$ is odd, $a_j(\bk)$ are constrained as
\begin{align}
    \text{Re}\left[a_j(\bk)\right] &= \begin{cases} -\text{Re}\left[a_j(-\bk)\right], &j \text{ even}, \\ \text{Re}\left[a_j(-\bk)\right], &j \text{ odd},\end{cases}
    \\
    \text{Im}\left[a_j(\bk)\right] &= \begin{cases} \text{Im}\left[a_j(-\bk)\right], &j \text{ even}, \\ -\text{Im}\left[a_j(-\bk)\right],&j \text{ odd}.\end{cases}
\end{align}
This means that the real and imaginary parts of $a_j(\bk)$ changes roles, and consequently so does  $\mathcal{T}$ and $\mathcal{C}$ symmetry in the topological classification.
Hence, $H_{\BR}(\bk)$ will be $\mathcal{T}$- and $\mathcal{C}$-symmetric with generators $C_+C_-$ and $\mathbb{I}$, respectively.
The topological classification is summarized in Table~\ref{tab:nonloc}, and coincides with that of $\mathcal{T}$ and $\mathcal{I}$ symmtery.

\subsection{Mayer-Vietoris arguments and vector bundle classification}
The symmetry-induced topological properties of EP$n$s of codimensions $2n-2$ are not as directly apparent in terms of cohomology groups.
Although the global charge-cancellation theorem, in terms of the symmetry-violating resultant winding number, follows, adopting the same method as in  Sec.~\ref{sec:NHMVS} will not provide any information related to the symmetry.
In particular, it will not result in a minimum number of four EP$n$s, which is a consequence of the non-local symmetries.
To encode this information, the cohomology reasoning has to be modified.
Earlier works have successfully provided such a cohomology description of the topological invariants appearing in (Hermitian) systems subject to $\mathcal{T}$ symmetry through the classification of quaternionic vector bundles (for $\mathcal{T}\mathcal{T}^*=-1$)~\cite{Nittis2014,Nittis2015}, ``real'' Bloch bundles (for $\mathcal{T}\mathcal{T}^*=+1$)~\cite{Nittis2018}, and equivariant and twisted cohomology and homology~\cite{Thiang2017}.
These studies, however, are restricted to low-dimensional cases, with four-dimensional systems being the most abstract systems treated.
Since a cohomology picture describing the symmetry-induced spectral topology of EP$n$s would be based on a general classification, it comprises, to our knowledge, an open mathematical research question.
It is therefore beyond the scope of the current manuscript and left for future works of an even stronger mathematical character.

\section{Non-Local Symmetries: Example Models and Physical Interpretations of the $\pi_d$-invariant} \label{sec:NLPhys}
To complement the abstract classification scheme developed in the previous section, this section is devoted to provide a physical picture and interpretation of the predicted topological invariants.
By using the mapping to the resultant Hamiltonian, Sec.~\ref{sec:TIs} provides the physical manifestation of the $\pi_d$-invariants describing the topology in the absence of EP$n$s. 
These claims are then strengthened by specifically considering the cases of two [Sec.~\ref{sec:twobands}], three [Sec.~\ref{sec:threebands}], four [Sec.~\ref{sec:fourbands}], and five [Sec.~\ref{sec:fivebands}] bands, respectively, before the fully general case is treated in Sec.~\ref{sec:genbands}.

\subsection{Topological invariants} \label{sec:TIs}
In Table~\ref{tab:nonloc}, both the $\pi_{d-1}$ and $\pi_d$-invariants are listed since they provide complementary information.
The $\pi_{d-1}$-invariant provides topological information more directly related to EP$n$s appearing at the high-symmetry points; by enclosing the EP$n$ by a symmetry-preserving sphere on which no EP$n$s exists, the topological invariant calculated on that sphere is assigned the topological invariant for the EP$n$.
Note that only the EP$n$s appearing at the high-symmetry points can be enclosed by a sphere preserving the symmetry; a symmetry-preserving surface enclosing an EP$n$ {\it away} from the high-symmetry points must also necessarily enclose the symmetry partner-EP$n$ (otherwise the symmetry is broken). 
The non-trivial $\pi_{d-1}$-invariants indicate that, in addition to the resultant winding number, there is an invariant further classifying EP$n$s appearing at the high-symmetry points in a manner similar to Hermitian Kramers degeneracies.
The $\mathbb{Z}_2$-invariants are reminiscent of the Fu-Kane-Mele (FKM) invariant and means that these EP$n$s further act as FKM monopoles~\cite{Thiang2017}.
The $\pi_{d-1}$-invariant is not uniquely given by $\mathbb{Z}_2$; systems subject to $\mathcal{T}$, $\mathcal{I}$ or $\mathcal{C}^{\dagger}$ it is instead given by $\mathbb{Z}$ $(n=2)$ or $2\mathbb{Z}$ $(n=4)$.
This indicate that these EP$n$s act as monopoles of $\mathbb{Z}$ and $2\mathbb{Z}$ flux.
When the $\pi_{d-1}$-invariant is trivial, it instead suggests that these additional invariants are not present, meaning that there are no additional monopole fluxes.

The $\pi_d$-invariants indicate that the EP$n$s are responsible for further interesting topological phenomena when they are gapped out. 
In terms of the resultant Hamiltonian, the eigenvalue topology of a non-Hermitian $n$-band model without EP$n$s can be understood in terms of the eigenvector topology of a fully gapped Hermitian system through the resultant Hamiltonian.
It has to be stressed that although the resultant Hamiltonian is fully gapped, the same conclusion cannot be drawn for the non-Hermitian parent system. 
A nowhere-vanishing resultant vector only means that the non-Hermitian system hosts no EP$n$s, while there is nothing forbidding the existence of EP$m$s for $m<n$.

The $\pi_d$-invariants exclusively take $\mathbb{Z}_2$-values.
To draw some intuition for what these $\mathbb{Z}_2$-invariants (and their absence) mean in terms of the non-Hermitian eigenvalue topology, it is again fruitful to resort to their interpretation in terms of Hermitian eigenvector topology.
There, the $\mathbb{Z}_2$-invariants can be related to what is known as Dirac strings~\cite{Thiang2017} and their topological stability. 
Dirac strings emerge between Hermitian nodal points and are the sources of their topological properties.
When nodal points are gapped out in specific ways with respect to the high-symmetry  points, the Dirac strings remain in the absence of the nodal points.
Although not present in non-Hermitian spectra, there is one feature whose properties are analogous to those of the Dirac strings, namely the bulk Fermi arcs~\cite{NHreview}.
Having topologically non-trivial Dirac strings means that they cannot be removed from the spectrum without passing through a gapless point.
The same holds true for bulk Fermi arcs; topologically non-trivial bulk Fermi arcs can only be removed from the spectrum by passing through an EP. 
Table~\ref{tab:nonloc} therefore tells us that $n$-band systems without EP$n$s host potentially topologically non-trivial bulk Fermi arcs on the surface of EP$(n-1)$s.
In other words, the $\mathbb{Z}_2$-invariant indicate that, when present, there exists two topologically different gapped phases, induced by how the EP$n$s are annihilated; it indicates the existence of an even or an odd number of non-trivial bulk Fermi arcs, and hence predicts the existence of a single topologically protected symmetry-induced bulk Fermi arc, which we refer to as a $\mathbb{Z}_2$-protected Fermi arc, in several $n$-band models without EP$n$s.

In what follows, we will show, both through simple toy models, but also through abstract general arguments, that the $\mathbb{Z}_2$-invariants physically manifest exactly as $\mathbb{Z}_2$-protected Fermi arcs.
These arguments will make it clear why topological bulk Fermi arc cannot emerge in systems with trivial $\pi_d$-invariant, and furthermore show that they indeed exist in the cases where the $\pi_d$-invariant is $\mathbb{Z}_2$.
The key lies in how the symmetry-induced spectral constraints make the eigenvalues ``mirror'' themselves or each other in the Brillouin zone.

\subsection{The case of two bands} \label{sec:twobands}
Since the $\pi_d$-invariant is predicted to be trivial for all non-local symmetries, it is sufficient to study only one example to illustrate how the bulk Fermi arc is gapped out.
Consider a two-band model described by
\begin{align}
    H(\bk) &= \begin{pmatrix} 0&1\\ a_R(\bk)+ia_I(\bk) &0  \end{pmatrix}, \label{eq:TRSdaggermodel}
    \\
    a_R(\bk) &=\cos k_x + \cos k_y-m_R,
    \\
    a_I(\bk) &= \cos k_y+\frac{1}{3}\cos^2 k_x -m_I.
\end{align}
\begin{figure*}
\includegraphics[width=\textwidth]{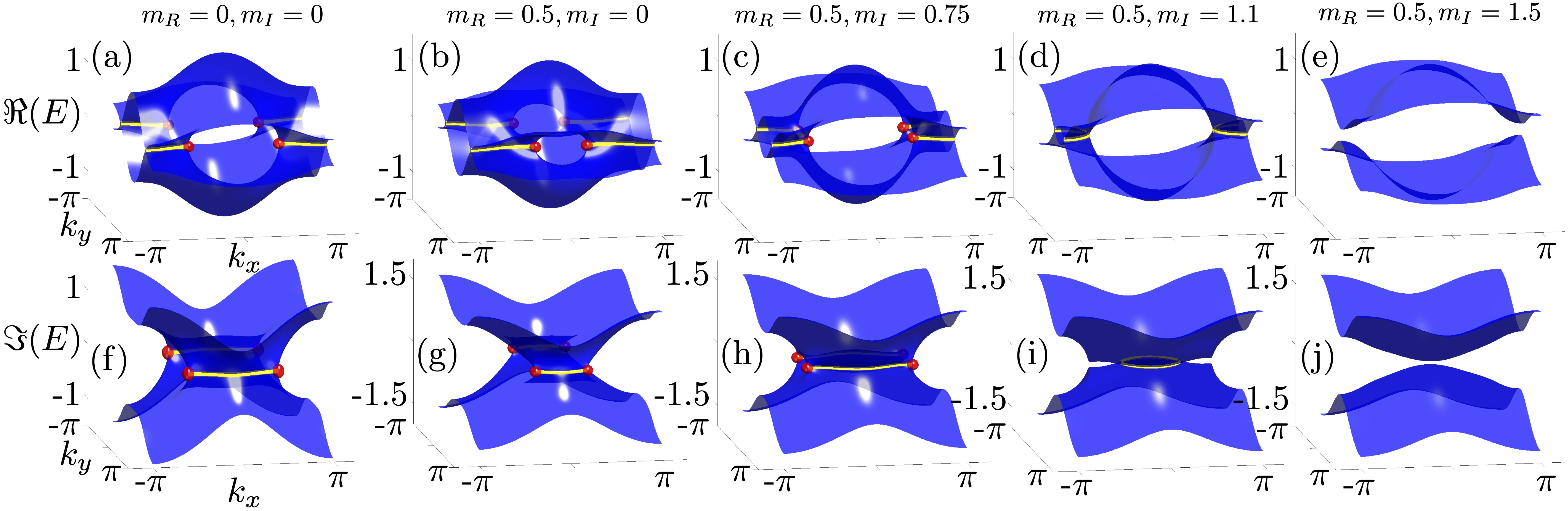}

\caption{Real and imaginary parts of the eigenvalues of the Hamiltonian given by Eq.~\eqref{eq:TRSdaggermodel} in panels (a)-(e) and (f)-(j), respectively. 
The red points illustrate $\mathcal{T}^{\dagger}$-symmetric EP2s, with their concomitant bulk Fermi and i-Fermi arcs highlighted with the yellow lines in (a)-(e) and (f)-(j), respectively. 
Varying the parameters $m_R$ and $m_I$ causes the EP2s to move in the Brillouin zone, shown in (a)-(c) and (f)-(h). 
Eventually, the EP2s overlap pairwise, causing them to annihilate and thus gapping out the spectrum. The annihilation of the EP2s does not cause the bulk Fermi and i-Fermi arcs to vanish, but rather to close, as shown in panels (d) and (i). 
The triviality of the second homotopy group for two-band models in two dimensions suggests that these are not topological, but can be removed from the spectrum without passing through a spectral phase including an EP2.
This is confirmed in panels (e) and (j), where increasing the value of $m_I$ causes the bulk Fermi arcs to vanish.
Consequently the $\mathcal{T}^{\dagger}$ and $\mathcal{P}$ symmetries do not induce any additional topological spectral features in two-dimensional two-band systems.
\label{fig:TBFA} }
\end{figure*}
The coefficients of the characteristic polynomial of $H(\bk)$ in Eq.~\eqref{eq:TRSdaggermodel} are even, and thus satisfy the constraints induced by $\mathcal{T}^{\dagger}$, $\mathcal{P}$, and $\mathcal{C}$ symmetry.
The corresponding real and imaginary spectrum are presented in Fig.~\ref{fig:TBFA} for different values of $m_R$ and $m_I$.
The symmetry constrains the minimum number of EP2s to be four, giving two open bulk Fermi arcs (both in real and imaginary energy).
Even when these EP2s are annihilated, the bulk Fermi arcs still persist, forming closed curves within the Brillouin torus.
Importantly, these can be removed entirely without the EP2s reappearing, meaning that the system does not have to pass through a gapless phase in order for the bulk Fermi arcs to vanish.

The reason for why the parity of the $a_I(\bk)$ (recall that $\mathcal{C}^{\dagger}$, $\mathcal{T}$ and $\mathcal{I}$ symmetry constrains $a_I(\bk)$ to be odd rather than even) doesn't matter can be understood from the expressions defining the Fermi arcs.
These are given by~\cite{ourknots},
\begin{equation}
\text{Re}\left[\lambda(\bk)\right] = 0 \iff \text{Im}\left[\lambda^2(\bk)\right]=0, \quad \text{Re}\left[\lambda^2(\bk)\right]\leq 0,
\end{equation}
with $E(\bk)$ denoting the eigenvalues of $H(\bk)$.
The gapping process can therefore be done completely independent of $a_I(\bk)$, and just by tuning $a_R(\bk)$ to reach a parameter regime where it is positive-definite, something that does not have to occur through an EP.

\subsection{The case of three bands}\label{sec:threebands}

\subsubsection{Time-reversal$^{\dagger}$ and Parity symmetry: Toy model}
\begin{figure*}
\includegraphics[width=\textwidth]{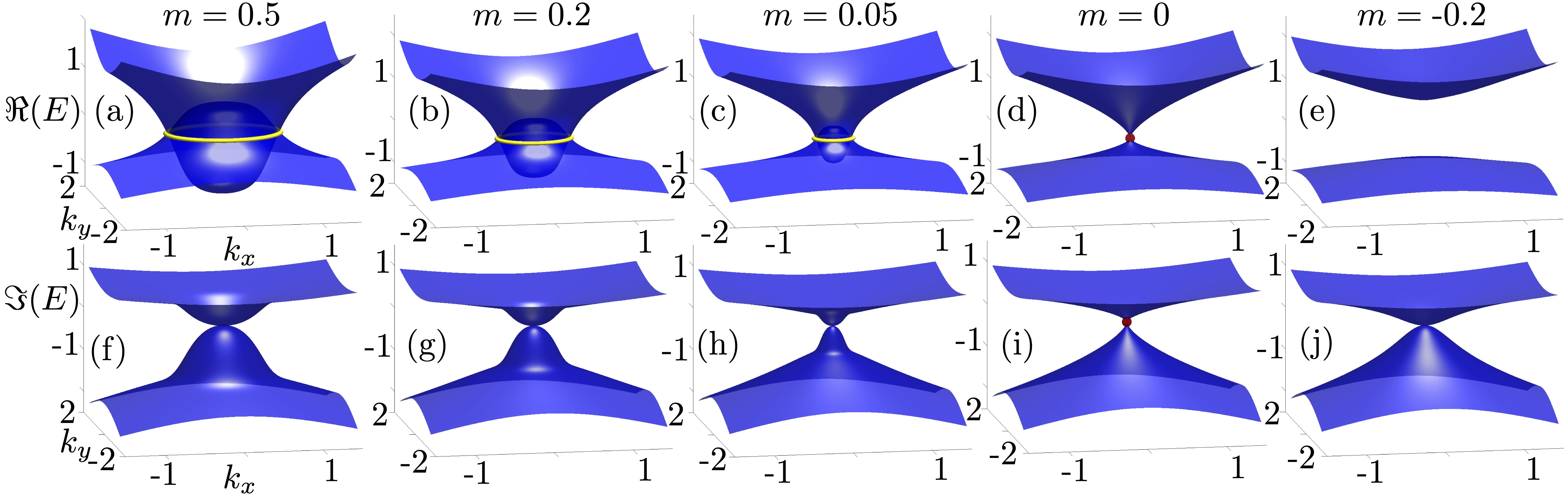}

\caption{Real and imaginary parts of the eigenvalues of the Hamiltonian given by Eq.~\eqref{eq:3BMod} in panels (a)-(e) and (f)-(j), respectively. 
The red points illustrate EP3s, with their concomitant bulk Fermi arcs highlighted with the yellow lines in (a)-(e). 
Varying the parameter $m$ causes the bulk Fermi arcs to change form within the Brillouin zone.
When $m$ changes sign, the bulk Fermi arcs either disappear (from positive to negative) or appear (from negative to positive).
This necessarily happens through a gapless point, i.e., an EP3, which then marks a topological phase transition point, something that did not happen in the two-dimensional case illustrated in Fig.~\ref{fig:TBFA}.
This is an example on how the non-trivial eigenvalue topology manifest in non-Hermitian systems subject to $\mathcal{T}^{\dagger}$ or $\mathcal{P}$ symmetry in four dimensions.
\label{fig:NTBFA} }
\end{figure*}

For the case of three bands, Table~\ref{tab:nonloc} indicates that the process illustrated in Fig.~\ref{fig:TBFA} is not always allowed---the classification frameworks predict the existence of $\mathbb{Z}_2$-protected Fermi arcs that can only disappear by passing through an EP3 in systems subject to $\mathcal{T}^{\dagger}$ and $\mathcal{P}$ symmetry.
This can be visualized by considering a model on the following form: 
\begin{equation} \label{eq:3BMod}
H(\bk) = \begin{pmatrix} 0&1&z_2(\bk)\\0&0&1\\z_1(\bk) &0&0 \end{pmatrix}.
\end{equation}
The unitary operator, $\mathcal{C}_+$, defining the time-reversal symmetry is
\begin{equation}
    \mathcal{C}_+ = \begin{pmatrix}
        0&0&1\\
        0&1&0\\
        1&0&0
    \end{pmatrix},
\end{equation}
and the characteristic polynomial of $H(\bk)$ reads,
\begin{equation}
P_3(\lambda; \bk) = -\lambda^3+z_1(\bk)z_2(\bk)\lambda+z_1(\bk).
\end{equation}
Here, $\bk$ denotes a four-dimensional momentum vector, and $z_{1,2}$ are continuously differentiable functions of $\bk$ taking complex values. 
The symmetry constraint forces $z_{1,2}(\bk) = z_{1,2}(-\bk)$.
The discriminant of this polynomial takes the form
\begin{equation}
D_{\lambda}(\bk) = -z_1(\bk)\left[27 z_1(\bk)+4z_2^3(\bk)\right],
\end{equation}
which indicates that $H(\bk)$ has degenerate eigenvalues when $z_1(\bk) =0$ or $27z_1(\bk) +4z_2^3(\bk) = 0$.
On the two-fold degenerate subspace defined by $4z_1(\bk)+27z_2^3(\bk)=0$ there will be arcs where all the eigenvalues have identical real parts---these define the bulk Fermi arcs that are claimed to source the topology indicated by Table~\ref{tab:nonloc}.
On this subspace, the eigenvalues read,
\begin{equation}
\lambda_{1,2}(\bk) = -\left[4z_1(\bk)\right]^{\frac{1}{3}}, \quad \lambda_3(\bk) = \left[\frac{z_1(\bk)}{2}\right]^{\frac{1}{3}}.
\end{equation}
Constraining the system to the degenerate surface corresponds to solving for two of the momentum components.
For the sake of illustration, we choose $z_1 = -k_x^2-k_y^2 + i\left(m - k_x^2-k_y^2\right)$ on the degenerate surface.
The real and imaginary parts of the corresponding eigenvalues are displayed in Fig.~\ref{fig:NTBFA}. By varying $m$, the bulk Fermi arcs changes accordingly, and are eventually gapped out when $m$ goes from being positive [Fig.~\ref{fig:NTBFA}(a)-(c) and (f)-(h)] to being negative [Fig.~\ref{fig:NTBFA} (e) and (j)].
Exactly when $m=0$ [Fig.~\ref{fig:NTBFA} (d) and (i)], the system hosts an EP3, marking the transition from the non-trivial to the trivial phase.
This is fundamentally different from the two-band system studied above, where the bulk Fermi arc was allowed to disappear without passing through a gapless point.
This lone bulk Fermi arc is the simplest example of a $\mathbb{Z}_2$-protected Fermi arc.

\subsubsection{The remaining symmetries: General arguments} \label{sec:threebandsgeneral}
Since all the remaining symmetries are classified within symmetry class BDI when $n=3$, their topological properties will be treated simultaneously.
The reasoning will however be carried out using the  spectral constraint induced by $\mathcal{C}$ symmetry,
\begin{equation}
    \left\{\lambda(\bk)\right\} = \left\{-\lambda(-\bk)\right\},
\end{equation}
keeping in mind that the other cases can be obtained through a change of basis.
The argument here will be of a more general kind than those carried out for the example models above.

\begin{figure*}[hbt!]
\includegraphics[width=\textwidth]{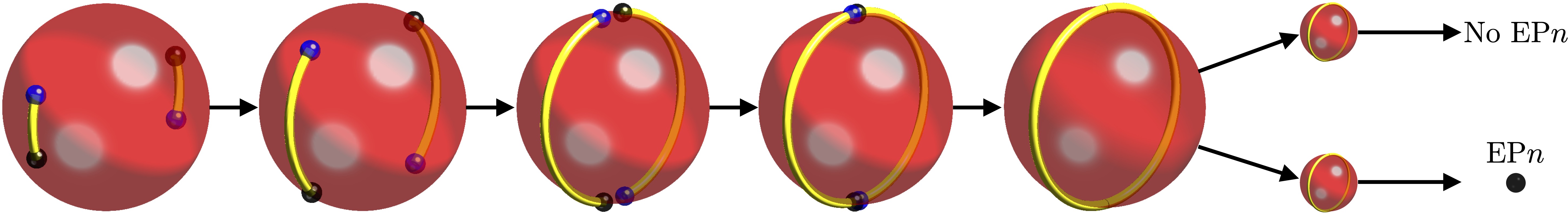}

\caption{An illustration of how the annihilation of EP$n$s (black and blue dots) on a symmetry-preserving sphere of EP$(n-1)$s (red sphere) potentially leads to topologically non-trivial bulk-Fermi arcs (yellow arcs). Here, the symmetries constrains the sphere to be centered around the origin of the Brillouin zone.
The bulk Fermi arcs connect EP$n$s of opposite resultant winding numbers. 
Since the underlying symmetry enforces the minimal number of EP$n$s to be four, an EP$n$s can annihilate by merging with another to whom it is not connected by a Fermi arc. 
This leads to a bulk Fermi arc symmetric on the sphere in the absence of EP$n$s. 
Whether the Fermi arc is topological or not depends on which eigenvalues coalesce to form the sphere of EP$(n-1)$s, the number of bands, and what symmetry is present, which is explained in Sec.~\ref{sec:NLPhys}.
\label{fig:EP3Ann} }
\end{figure*}

The symmetry-induced spectral relations can be imposed in several ways, but Table~\ref{tab:nonloc} predict that the systems remain topologically trivial regardless of this choice. 
Start with the case where the symmetry-induced spectral relations take the form
\begin{equation} \label{eq:oddEVs}
    \lambda_j(\bk)=-\lambda_j(-\bk), \quad j=1,2,3.
\end{equation}
Such a system may host two-dimensional surfaces of EP2s and point-like EP3s located on the surfaces of EP2s.
The underlying symmetry enforces the surfaces of EP2s to appear in a symmetric fashion in the Brillouin torus; any surface of EP2s need to have a symmetric partner at opposite momentum.
However, there is nothing forbidding the EP2 surface to be symmetric in itself.
Assume therefore that there exists EP2s forming one symmetry-preserving sphere around a high-symmetry point as
\begin{equation}
    S^2_{\text{EP2}} = \left\{ \bk : \lambda_1(\bk)=\lambda_2(\bk)\right\}.
\end{equation}
Assume further that there exists two pairs of EP3s on this sphere of EP2s, which are connected by a bulk Fermi arc.
The EP3s can be merged pairwise [cf. Fig.~\ref{fig:EP3Ann} where such a process is schematically depicted] and will hence result in a bulk Fermi arc along the equator.
When shrinking the size of the sphere of EP2s, it will eventually reach a critical point at the origin of momentum space.
Since all eigenvalues are odd functions of momentum, they all necessarily vanish when $\bk=0$, making it seem as if the system necessarily pass through an EP3 when the bulk Fermi arc is gapped out.
Although the origin marks a threefold eigenvalue degeneracy, this does not comprise a stable EP3; the coefficients of the characteristic polynomial do not all necessarily vanish at these points since only half of them are odd in $\bk$ [cf., e.g., Eqs.~\eqref{eq:charpolPHS1} and \eqref{eq:charpolPHS2}].
This means that the threefold eigenvalue degeneracy can be gapped out by an infinitesimally small symmetry-preserving perturbation, which furthermore breaks Eq~\eqref{eq:oddEVs}; having all eigenvalues being odd in $\bk$ is not a stable symmetry-induced spectral constraint.
Therefore, the corresponding bulk Fermi arcs can be gapped out without passing through an EP3, confirming that they are trivial.

Now, choose instead the spectral relations to be, e.g., 
\begin{equation}
    \lambda_1(\bk)=-\lambda_1(-\bk), \quad \lambda_2(\bk)=-\lambda_3(-\bk),
\end{equation}
and consider a sphere of EP2s made up by 
\begin{align}
    \left(S^2_{\text{EP2}}\right)^+ &=\left\{ \bk : \lambda_1(\bk)=\lambda_2(\bk)\right\},
    \\
    \left(S^2_{\text{EP2}}\right)^- &=\left\{ \bk : \lambda_1(\bk)=\lambda_3(\bk)\right\},
\end{align} 
where $+(-)$ denotes the upper (lower) hemisphere.
Extra care has to be taken on the equator though.
Parameterize the equation as a circle by some angle $\phi$. 
Then, letting 
\begin{equation}
\lambda_1(\bk)=\lambda_2(\bk), \quad \phi\in [0,\pi),
\end{equation}
the symmetry induces that 
\begin{equation}
\lambda_1(\bk)=\lambda_3(\bk), \quad \phi\in [\pi,2\pi).
\end{equation}
The fact that $\lambda_1$ is odd means that its value on the upper hemisphere must be reflected with a sign in the lower hemisphere.
Hence, $\lambda_1$ need to cross a zero somewhere on the sphere of EP2s, meaning that for some $\bk=\bk_0$, $\lambda_1(\bk_0)=\lambda_2(\bk_0)=0$. 
But since $\lambda_2(\bk)=-\lambda_3(\bk)$, also $\lambda_3(\bk_0)=0$, and since this threefold eigenvalue degeneracy necessarily occur at non-trivial $\bk=\bk_0$, this point comprise an EP3.
Therefore, in such a system the EP3 cannot be completely gapped out, meaning that the information stemming from the $\pi_d$-invariant is not applicable.
This explains why the $\pi_d$-invariant is predicted to be trivial for $n=3$.

\subsection{The case of four bands} \label{sec:fourbands}
The classification scheme predicts the existence of a non-trivial $\mathbb{Z}_2$-valued topological invariant in non-Hermitian $\mathcal{C}$-symmetric four band-models.
Interestingly enough, this case can be understood directly from the behavior of a degree-four polynomial.
Consider a polynomial on the form
\begin{equation}
    P_4(\lambda) = \lambda^4-a_2\lambda^2-a_1\lambda-a_0,
\end{equation}
where the $\bk$-dependence has been neglected for brevity, and will be re-introduced when appropriate.
$P_4(\lambda)$ has a triple root (corresponding to EP3s in our physical models) when the discriminant vanishes, and, additionally, the quantity
\begin{equation} \label{eq:delta0=0}
    \Delta_0 = a_2^2-12a_0 = 0,
\end{equation}
for non-zero $a_2$ and $a_0$.
The triple root of $P_4$ is furthermore a root of its second derivative, $P''_4$, which means that it is also the root of the remainder $r(\lambda)$ obtained when dividing $P_4$ by $P_4''$.
The Euclidean algorithm yields,
\begin{equation}
    P_4(\lambda) = P''_4(\lambda) \left(-\frac{\lambda^2}{12}-\frac{5a_2}{72}\right)-a_1\lambda-a_0-\frac{10 c^2}{72}.
\end{equation}
Therefore, the triple root can be written as,
\begin{equation}
    \lambda_0 = -\frac{1}{a_1}\left(a_0+\frac{5a_2^2}{36}\right).
\end{equation}
This means that three of the eigenvalues are given by $\lambda_0$ at the EP3s.
The remaining eigenvalue is furthermore given by,
\begin{equation}
    \lambda_4 = -3\lambda_{1,2,3} = -3\lambda_0,
\end{equation}
which follows from the trace of the parent non-Hermitian matrix being zero.
Using Eq.~\eqref{eq:delta0=0}, this simplifies further as,
\begin{equation}
    \lambda_0 = -\frac{2}{9} \frac{a_2^2}{a_1}.
\end{equation}

Now, when $\Delta_0=0$, the discriminant of $P_4$ takes the form,
\begin{equation}
    D_4 = -\left(\frac{4a_2^2}{3}\right)^3+a_2^3\left(4a_1\right)^2-27a_1^4,
\end{equation}
and solving $D_4=0$ yields,
\begin{equation}
    a_2^3 = \frac{27}{8}a_1^2,
\end{equation}
which means that the triple root can be written as 
\begin{equation}
    \lambda_0 = \pm\sqrt{\frac{1}{6} a_2}:= \pm \sqrt{\tilde{a}_2}.
\end{equation}
Now, the difference between the symmetries can be understood directly from the induced spectral constraints.
For the sake of clarity, the symmetries yielding different topological classifications will be treated separately.
\subsubsection{Particle-hole symmetry}
The $\mathcal{C}$ symmetry-induced spectral relation reads $\left\{\lambda(\bk)\right\} = \left\{-\lambda(-\bk)\right\}$.
Thus, assuming that the threefold eigenvalue degeneracy discussed above takes the form of a sphere centered around the origin, all four eigenvalues coalesce with value $0$ when the radius of the EP3 sphere is taken to zero.
Consequently, $a_2(0)=0$ necessarily.
Since $a_1(\bk)=-a_1(-\bk)$, it also vanishes at the origin.
To make sure that Eq.~\eqref{eq:delta0=0} still holds, also $a_0(0)=0$.
This means that all coefficients of the characteristic polynomial vanish at the origin when shrinking a sphere of EP3s, and thus a bulk Fermi arc located on this EP3 sphere can only be gapped out by passing through an EP4, which means that it is topologically non-trivial---it is a $\mathbb{Z}_2$-protected Fermi arc.
This explains why the $\pi_d$-invariant for $\mathcal{C}$-symmetric systems is $\mathbb{Z}_2$ for $n=4$.

\begin{figure*}
\includegraphics[width=\textwidth]{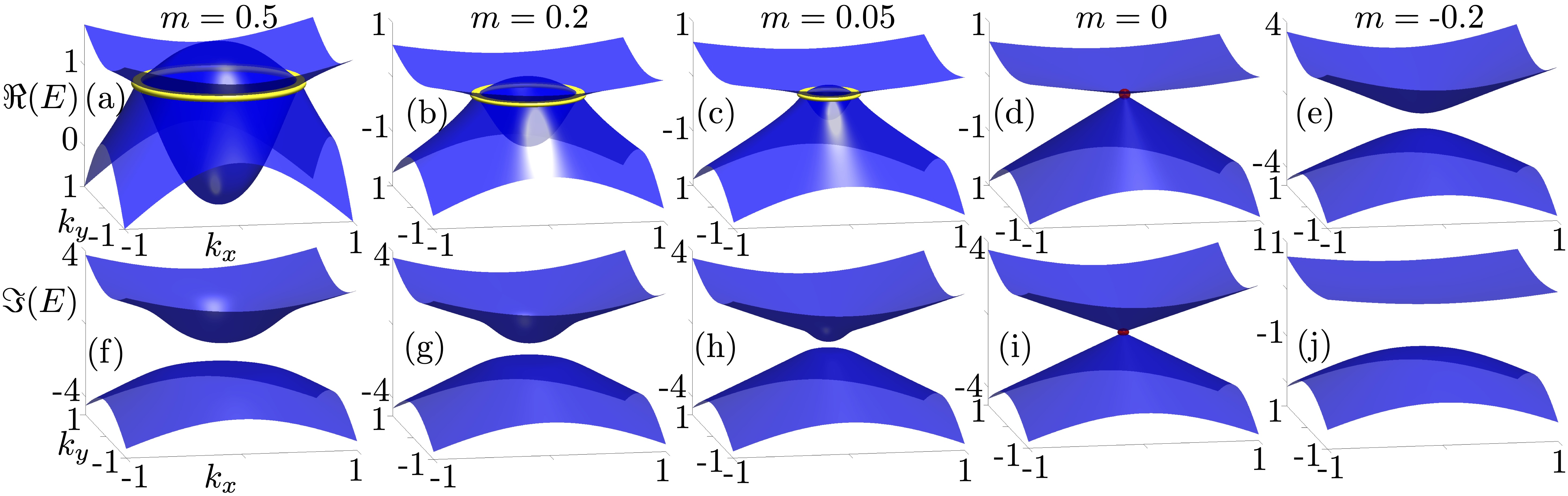}

\caption{Real and imaginary parts of the eigenvalues of the Hamiltonian given by Eqs.~\eqref{eq:examPHS} and \eqref{eq:a2PHS} constrained to a sphere of EP3s centered around the origin of momentum space in panels (a)-(e) and (f)-(j), respectively.
The red points illustrate EP3s, with their concomitant bulk Fermi arcs highlighted with the yellow lines in (a)-(e). 
Varying the parameter $m$ causes the bulk Fermi arcs to change form within the Brillouin zone.
When $m$ changes sign, the bulk Fermi arcs either disappear (from positive to negative) or appear (from negative to positive).
This necessarily happens through a gapless point, i.e., an EP4, which then shows that a topological phase transition happen also in four band systems subject to $\mathcal{C}$ symmetry in six dimensions; this feature is not present for any other symmetry when the number of bands is even.
\label{fig:NTBFAPHS} }
\end{figure*}

This feature can be specifically illustrated by considering a toy model on the form,
\begin{equation} \label{eq:examPHS}
H_{\mathcal{C}}(\bk) = \begin{pmatrix} 0&1&0&0\\ \frac{c(-\bk)}{2}&0&1&0\\ -\frac{b(-\bk)}{2} & 0&0&1\\ a_0(\bk)+c(\bk) c(-\bk) & \frac{b(\bk)}{2} & \frac{c(\bk)}{2} & 0 \end{pmatrix},
\end{equation}
which satisfies $\mathcal{C}$ symmetry with generator
\begin{equation}
\mathcal{T}_- = \begin{pmatrix} 0&0&0&i\\0&0&-i&0\\0&i&0&0\\-i&0&0&0 \end{pmatrix}.
\end{equation}
Here, $\bk$ denotes a six dimensional momentum vector.
The corresponding characteristic polynomial reads,
\begin{equation}
P_4(\lambda) = \lambda^4 - \left[\frac{c(\bk)+c(-\bk)}{2}\right] \lambda^2 - \left[\frac{b(\bk)-b(-\bk)}{2}\right] - a_0(\bk).
\end{equation}
The extra term $c(\bk)c(-\bk)$ in the lower left corner does not alter the reasoning, but it simplifies the characteristic polynomial and is therefore placed there for convenience.
Assuming that three of the eigenvalues form an EP3 sphere centered around the origin, the eigenvalues read,
\begin{equation}
\lambda_{1,2,3}(\bk) = \pm \sqrt{\frac{a_2(\bk)}{6}}, \quad \lambda_4(\bk) = \pm \sqrt{\frac{3 a_2(\bk)}{2}},
\end{equation}
with $a_2(\bk):= \frac{c(\bk)+c(-\bk)}{2}$.
Now, assume that a sphere of EP3s centered around the origin is formed by tuning 4 of the 6 momentum components.
The eigenvalues constrained to this EP3 sphere can therefore be described as a two-dimensional problem.
The spectrum, constrained to the sphere of EP3s, is presented in Fig.~\ref{fig:NTBFAPHS} with the specific choice,
\begin{equation} \label{eq:a2PHS}
a_2(\bk) = -k_x^2-k_y^2+i\left(k_x^2+k_y^2-m\right),
\end{equation}
for various choices of $m\in \mathbb{R}$.
Note that this is the form of $a_2(\bk)$ when constrained to the sphere of EP3s, and hence it does not pass through the origin of momentum space unless $m=0$.
Hence, $m$ is directly related to the radius of the sphere of EP3s.
When $m>0$, a bulk Fermi arc is formed, cf. Fig.~\ref{fig:NTBFAPHS} (a)-(c), which is gapped out as soon as $m<0$, cf. Fig.~\ref{fig:NTBFAPHS} (e).
But this transition necessarily happens through an EP4 emerging at $m=0$, cf. Fig.~\ref{fig:NTBFAPHS} (d) and (i), meaning that the bulk Fermi arc is indeed $\mathbb{Z}_2$-protected.

\subsubsection{The remaining symmetries}
The symmetry-induced spectral constraints for the remaining symmetries does not constrain the fourth eigenvalue to any specific value at the origin.
Therefore, the eigenvalue not being part of the EP3 sphere doesn't necessarily take the value zero at the origin.
It can in fact take any value ($\mathcal{T}^{\dagger}$/$\mathcal{P}$ symmetry), any real value ($\mathcal{T}/\mathcal{I}$ symmetry), or any imaginary value ($\mathcal{C}^{\dagger}$ symmetry), and thus it isn't constrained to coalesce with the other eigenvalues when the radius of the EP3 sphere is taken to zero.
Hence, a bulk Fermi arc located on such an EP3 sphere may be removed from the spectrum without passing through an EP4, which means that it is trivial.
This explains the trivial $\pi_d$-invariant for these symmetries when $n=4$.

\subsection{The case of five bands} \label{sec:fivebands}
Table~\ref{tab:nonloc} indicates that all symmetries give rise to a gapped topological phase in five band systems, classified by a $\mathbb{Z}_2$-invariant.
In the cases of $\mathcal{T}^{\dagger}$ and $\mathcal{P}$ symmetry, this invariant will be discussed in the following subsection.
Here, we focus on the remaining symmetries simultaneously, as they all belong to the BDI class.
Again, the reasoning will be carried out specifically for the spectral constraint given by $\left\{\lambda(\bk)\right\} = \left\{-\lambda(-\bk)\right\}$.

Assume that the symmetry is reflected in spectral constraints on the form $\lambda_j(\bk)=-\lambda_j(-\bk)$ for $j=1,2,3$, and $\lambda_4(\bk)=-\lambda_5(-\bk)$.
Let then furthermore a sphere of EP4s be made up by
\begin{equation}
    S^2_{\text{EP4}}  = \left\{ \bk : \lambda_1(\bk)=\lambda_2(\bk)=\lambda_4(\bk)=\lambda_5(\bk)\right\}.
\end{equation}
There will be a reflection of both $\lambda_1$ and $\lambda_2$ on opposite hemispheres, but since $\lambda_3$ is not involved in the EP4-sphere at all, these reflections will not induce an EP5 [this situation should be compared to the three-band case within the BDI-class, treated in Sec.~\ref{sec:threebands}, where EP3s on a sphere of EP2s could not be annihilated].
Therefore EP5s can be created, moved, and annihilated---all in a symmetric fashion.
Such a process will form a bulk Fermi arc.
Shrinking the radius of the EP4 sphere to 0 will result in a symmetry-induced EP5 at the origin. 
To see that this will indeed result in an EP5, we again need to look at the coefficients of the characteristic polynomial, given by Eqs.~\eqref{eq:charpolPHS1} and \eqref{eq:charpolPHS2}.
At the origin, the characteristic polynomial takes the form,
\begin{equation}
    P_{\mathcal{C}}(\lambda,0) = -\lambda^5+a_2(0)\lambda^2 + a_0(0),
\end{equation}
since $a_{0,2}(\bk) = a_{0,2}(-\bk)$ and $a_{1,3}(\bk) = -a_{1,3}(-\bk)$.
The corresponding eigenvalues are
\begin{align}
    \lambda_3(0) &= 0, 
    \\
    \lambda_{1,2,4,5}(0) &= \pm_1 \frac{ \sqrt{a_0(0)\pm_2 \sqrt{a_0^2(0)+4a_2(0)}}}{\sqrt{2}},
\end{align}
where $\pm_1$ and $\pm_2$ are independent of each other, and the four different components define the four remaining eigenvalues (the ordering is unimportant).
Since the origin is surrounded by a sphere of EP4s, the four eigenvalues that are degenerate on the sphere remain to be so at the origin.
Having $\lambda_{1,2,4,5}$ all coalescing at the origin forces $a_0(0)=a_2(0)=0$.
Hence, shrinking the sphere of EP4s to a point indeed results in an EP5, meaning that the bulk Fermi arc can only be removed by passing through a phase including an EP5.
This explains the non-trivial $\pi_d$-invariant for $n=5$.

To understand the generalization to an arbitrary number of bands is provided by Bott periodicity in $d$; the same topology is expected as $d\to d+8$, which means that identical outcomes are expected when $n\to n+4$.

\subsection{Schematic arguments for the general cases} \label{sec:genbands}
Finally, this section outlines how to find the physical manifestation of the $\mathbb{Z}_2$-valued $\pi_d$-invariants in systems with an arbitrary number of bands.
The procedure will be performed exclusively for systems subject to $\mathcal{T}^{\dagger}$ or $\mathcal{P}$ symmetry, but can be modified in a completely analogues manner to reproduce the general cases also for the remaining symmetries.

The feature illustrated in Fig.~\ref{fig:NTBFA} is not unique to three-band systems subject to the symmetry-induced spectral constraint $\left\{\lambda(\bk)\right\} = \left\{-\lambda(-\bk)\right\}$, but is rather expected generically in $\mathcal{T}^{\dagger}$- and $\mathcal{P}$-symmetric systems with an odd number of (at least three) bands.
The $\mathbb{Z}_2$-invariant expected in these systems can be understood in terms of a schematic argument involving the bands rather than looking at specific models.
To see this, consider again a three-band system in a four-dimensional momentum space.
Assume that the symmetry enforces spectral relations between the three eigenvalues as 
\begin{equation}
    \lambda_1(\bk)=\lambda_1(-\bk), \quad \lambda_2(\bk) = \lambda_3(-\bk).
\end{equation}
Assume further that the system hosts a sphere of EP2s located symmetrically around the high-symmetry point.
Consider first the case when the sphere of EP2s corresponds to a region in momentum space where 
\begin{equation}
S^2_{\text{EP2}} = \left\{\bk : \lambda_2(\bk)=\lambda_3(\bk)\right\}
\end{equation}
with four symmetry-induced EP3s appearing at the origin of momentum space.
The EP3s can be annihilated to form a bulk Fermi arc corresponding to the equator of the sphere, cf. Fig~\ref{fig:EP3Ann}.
When decreasing the radius of the sphere, the bulk Fermi arcs will shrink accordingly.
When the sphere is shrunk all the way to a point, the spectral constraints are still fulfilled, and the EP2 surface can be completely gapped out without passing through an EP3.
Consequently, the bulk Fermi arcs are also removed without passing through an EP3, indicating that it is topologically trivial---the symmetry has to be reflected differently in the spectral constraints in order for the existence of a $\mathbb{Z}_2$-protected Fermi arc to be made possible.

Second, consider the case when the sphere of EP2s is instead made up partly by a region where $\lambda_1(\bk)=\lambda_2(\bk)$, and partly by a region where $\lambda_1(\bk)=\lambda_3(\bk)$.
Note that whenever $\lambda_1(\bk)=\lambda_2(\bk)$, the symmetry induces $\lambda_1(-\bk)=\lambda_3(-\bk)$.
This means that the upper and lower hemispheres are respectively defined as 
\begin{align}
\left(S^2_{\text{EP2}}\right)^+ &= \left\{ \bk : \lambda_1(\bk)=\lambda_2(\bk)\right\},
\\
\left(S^2_{\text{EP2}}\right)^-&= \left\{ \bk : \lambda_1(\bk)=\lambda_3(\bk)\right\}.
\end{align}
Further, as in Sec.~\ref{sec:threebandsgeneral}, choose half the equator linking the two hemispheres to be $\lambda_1=\lambda_2$ and the other half to be $\lambda_1=\lambda_3$ in a fashion consistent with the symmetry.
Assume now that on the sphere, there are four symmetry-induced EP3s.
Naturally, if $\lambda_1(\bk)=\lambda_2(\bk)$, then the symmetry enforces $\lambda_1(-\bk)=\lambda_3(\bk)$, but the latter does not result in any additional surfaces of EP2s (it still corresponds to a points on the same sphere of EP2s).
The EP3s can annihilate in the same fashion as before, forming a bulk Fermi arc as the equator of the sphere.
When the radius of the sphere is decreased, the bulk Fermi arc shrinks accordingly.
When the sphere turns to a point, the symmetry now enforces all three eigenvalues to coalesce, as a consequence of how the EP2 sphere was originally constructed. 
Hence, the sphere of EP2s becomes a symmetry-induced EP3.
This means that the bulk Fermi arc cannot be annihilated from the spectrum without passing through a phase where the system hosts an EP3, and consequently, it is a $\mathbb{Z}_2$-protected Fermi arc.
This procedure is illustrated in Fig.~\ref{fig:EP3Ann}.

This argument can be straightforwardly generalized to an arbitrary number of eigenvalues. 
If $n$ is an even number, there is no way to construct a sphere of EP$(n-1)$s such that the spectral constraints enforce it to induce an EP$n$ when the radius is taken to 0.
If $n$ is an odd number, the sphere of EP$(n-1)$s is constructed in a way analogous to the one above.
With spectral constraints 
\begin{align}
    \lambda_1(\bk)&=\lambda_1(-\bk), \quad \lambda_{j}(\bk)=\lambda_{j+1}(-\bk), 
    \\
    j&\in \{2,...,n-1\}, \nonumber
\end{align}
the sphere should be partly made up by, e.g., a region where $\lambda_1=\lambda_2=...=\lambda_{n-1}$, and partly by a region where $\lambda_1=...=\lambda_{n-2}=\lambda_n$.
Annihilating the EP$n$s such that the bulk Fermi arc forms the equator of the sphere, decreasing its radius will eventually induce an EP$n$, making the bulk Fermi arcs topological.
The illustration provided in Fig.~\ref{fig:EP3Ann} applies also to this general case with the sphere corresponding exactly to the sphere of EP$(n-1)$s.


\section{Corollaries} \label{sec:COR}
The above classification scheme applies not only to point-like EP$n$s but also extends to other structures. In Sec.~\ref{sec:NDEP} the case of the so-called non-defective EPs is discussed, while Sec.~\ref{sec:ELines} is devoted to consequences in systems hosting EP$n$s of higher dimensions. Sec.~\ref{sec:gaptop} discusses topological properties of systems in the absence of EP$n$s.
\subsection{Non-defective EPs} 
\label{sec:NDEP}
\begin{table*}
\begin{tabular}{|c|c|c|c|}
\hline
{\bf NDEP$n$ Type} & {\bf Generic} &{\bf Pseudo-Hermitian}& {\bf Self Skew-Similar}\\
 \hline
 {\bf Codimension} & $2n^2-2$ & $n^2-1$&$2\lfloor \frac{n^2}{2} \rfloor-1$\\
 \hline
\multirow{2}{6em}{{\bf Hermitian Operator}}& \multirow{2}{8em}{$M = \sum_{i=1}^{2n^2-2}v_i \gamma^i$ $\{M,\gamma^{2n^2-1}\}=0$} &\multirow{2}{9em}{$M = \sum_{i=1}^{n^2-1}v_i \gamma^i$ $\{M,\gamma^{n^2}\}=0$, $n$ odd} &\multirow{2}{9em}{$M = \sum_{i=1}^{2\lfloor \frac{n^2}{2} \rfloor-1}v_i \gamma^i$ $\{M,\gamma^{2\lfloor \frac{n^2}{2} \rfloor}\}=0$} \\
  & & & \\
  \hline
    \multirow{2}{8em}{{\bf Corresponding Hermitian Deg.}}& \multirow{2}{6em}{$(2^{n^2-1})$-fold protected} &\multirow{2}{12em}{$2^{\frac{n^2-1}{2}}$-fold, prot., $n$ odd. $2^{\frac{n^2-2}{2}}$-fold gen., $n$ even} &\multirow{2}{7em}{$2^{\lfloor\frac{n^2}{2}\rfloor-1}$-fold protected} \\
  & & & \\
  \hline
   \multirow{2}{7em}{{\bf Topological Invariant}}& \multirow{2}{6em}{Class AIII, $\mathbb{Z}$} &\multirow{2}{10em}{Class A, $n$ even, $\mathbb{Z}$. Class AIII, $n$ odd, $\mathbb{Z}$.} &\multirow{2}{6em}{Class AIII, $\mathbb{Z}$} \\
  & & & \\
  \hline
  \multirow{2}{8em}{{\bf Vector Bundle}}& \multirow{2}{10em}{$T\mathbb{T}^{2n^2-2}$, rank $2n^2-2$} &\multirow{2}{10em}{$T\mathbb{T}^{n^2-1}$, rank $n^2-1$} &\multirow{2}{13em}{$T\mathbb{T}^{2\lfloor \frac{n^2}{2}\rfloor-1}$, rank $2\lfloor \frac{n^2}{2} \rfloor-1$} \\
  & & & \\
  \hline
  {\bf Vector Field} &$\mathbf{V} \in \Gamma(T\mathbb{T}^{2n^2-2})$ &$\mathbf{V} \in \Gamma(T\mathbb{T}^{n^2-1})$&$\mathbf{V} \in \Gamma(T\mathbb{T}^{n^2-1/n^2-2})$ \\
  \hline
\end{tabular}
 \caption{A summary of the topological classification of non-defective EP$n$s, including their codimension, relation to Hermitian degeneracies and symmetry classes, and their corresponding vector bundle interpretation. 
 This classification is obtained in a manner equivalent to the one used for EP$n$s, the difference here being the codimensions and thus, the rank of the vector bundle classification. 
 This provides the full topological picture of generic and similarity-protected non-defective EP$n$s, from the abstract notion of vector bundles to concrete Bloch Hamiltonians via the notion of sections of the vector bundle (or, equivalently, vector fields on the base manifold). The explicit dependence on the parameter $\bk$ has been neglected for brevity, and $\lfloor x \rfloor$ denotes the floor function of $x$.} \label{tab:sumresND}
\end{table*}

In addition to the previously studied EP$n$s, there are eigenvalue degeneracies corresponding to points where the parent matrix vanishes (or, more correctly, is proportional to the identity matrix). 
These have obtained several different names in the literature based on the situation in which they appear, all based on their similarity to Hermitian degeneracies. 
Some examples are Dirac points~\cite{NHTBT} and diabolic points~\cite{Keck2003,Xue2020,Wiersig2022,Zhang2023}. 
Here, they will be referred to as non-defective EPs, as in Ref.~\cite{Sayyad23}. 
This name is chosen based on that these points seem to be non-defective locally, in the sense that the Hamiltonian is diagonal at the point, but there exists no infinitesimally small neighborhood of the point where the system remains stable and diagonalizable.

Following a similar reasoning as for EP$n$s, it becomes clear that the non-defective EPs are bound to obey a separate doubling theorem, meaning that they are topological. 
This reasoning is applicable both for generic systems and those subject to similarities. 
For some non-Hermitian matrix
\begin{equation}
    M = \bd \cdot \boldsymbol{\gamma},
\end{equation}
with $\bd:\mathbb{T}^x \to \mathbb{C}^y$, and $\boldsymbol{\gamma}$ the vector of $\gamma$-matrices in the appropriate dimensions, non-defective EPs correspond to points where 
\begin{equation}
    \text{Re}\left(\bd\right) = \text{Im}\left(\bd\right) = 0.
\end{equation}
Introduce a new (real) vector as
\begin{equation}
    \mathbf{V} = \left[\text{Re}(d_1),...,\text{Re}(d_x),\text{Im}(d_1),...,\text{Im}(d_x)\right]^T,
\end{equation}
and from the vector $\mathbf{V}$, a Hermitian matrix can be defined in a way similar to how the Resultant Hamiltonian in Sec.~\ref{sec:TNEP} is defined,
\begin{equation}
    M = \sum_{i=1}^{x}v_i\gamma^i,
\end{equation}
with $v_i$ denoting the components of $\mathbf{V}$.
As for the eigenvalues of the resultant Hamiltonian, the eigenvalues of $M$ will display a generalized ``spin'' degeneracy.
The winding number of the vector $\mathbf{V}$ around the origin will give a topological invariant for the non-defective EP. 
In generic systems, this will amount to a winding of $(2n^2-3)$-spheres, while in pseudo-Hermitian systems it will be a winding number of $(n^2-2)$-spheres. 
For self skew-similar systems, the nature of the winding number will depend on the parity of $n$. 
For $n$ even, the winding will be around $(n^2-2)$-spheres, and for $n$ odd it will be around $(n^2-3)$-spheres. 
In summary, the corresponding winding numbers will be given by
\begin{align}
    W &\propto \oint_{S^x} (\mathbf{v}^{-1} d\mathbf{v})^{x}, \nonumber
    \\
    x&=\begin{cases} 2n^2-3, &\text{ generically,} \\
    n^2-2, &\text{ pseudo-Hermiticity,} \\
    \begin{cases} n^2-2, &n \text{ even,} \\ n^2-3, &n \text{ odd,} \end{cases} &\text{ self skew-similarity,}\end{cases}
\end{align}
where $\mathbf{v} = \frac{\mathbf{V}}{|\mathbf{V}|}$. The topological nature of these non-defective EPs is summarized in Table~\ref{tab:sumresND}, where also the corresponding vector bundle classification is listed (note that this follows directly from the reasoning applied to EP$n$s in Sec.~\ref{sec:VB}).

\subsection{EP$n$s in higher dimensions} \label{sec:ELines}
Although the above classification scheme revolves around $0$-dimensional EP$n$s, it can be directly generalized to EP$n$s of higher dimensions.
In other words, the scheme is not limited to EP$n$s in $2n-2$ (for generic systems and systems subject to non-local symmetries), $n-1$ (for systems subject to pseudo-Hermitian similarity or odd-band models subject to self skew-similarity) or $n$ (for even-band systems subject to self skew-similarity) dimensions, but extends to any higher dimensional space as long as the EP$n$s comprise a coalescence of all eigenvalues.
An intuitive picture of how this can be done can be extracted from the case of EP2s appearing as closed curves in generic systems in three dimensions. 
A curve of EP2s in three dimensions can potentially host both non-trivial $\pi_1$ and $\pi_2$-invariants. 
For instance, it may carry a non-trivial Chern number due to the eigenvector topology. 
However, the eigenvalue topology has a trivial invariant $\pi_2$.
This can be illustrated by considering a sphere enclosing the curve of EP2s. 
This induces natural map from the sphere to the the normalized resultant vector, which will be homotopy equivalent to a circle. 
Such maps are classified by $\pi_2(S^1)$, which is known to be trivial.
Consequently, the topological properties of EP2s in any dimension are directly determined by the behavior of the resultant vector along a circle winding around the manifold of EP2s. In other words, the topology is characterized by a map between circles in the base space and circles in the space of resultant vectors.

Similarly, the Abelian eigenvalue topology of EP$n$s is generally described by the maps between spheres in the base space and spheres in the space of resultant vectors.
The dimension of the spheres is $2n-3$ for generic EP$n$, $n-2$ for EP$n$ protected by pseudo-Hermitian similarity or self skew-similarity (for odd values of $n$), or $n-1$ for EP$n$s protected by self skew-similarity (for even values of $n$).

\subsection{Topology of systems without EP$n$s} \label{sec:gaptop}

\begin{figure*}
\includegraphics[width=\textwidth]{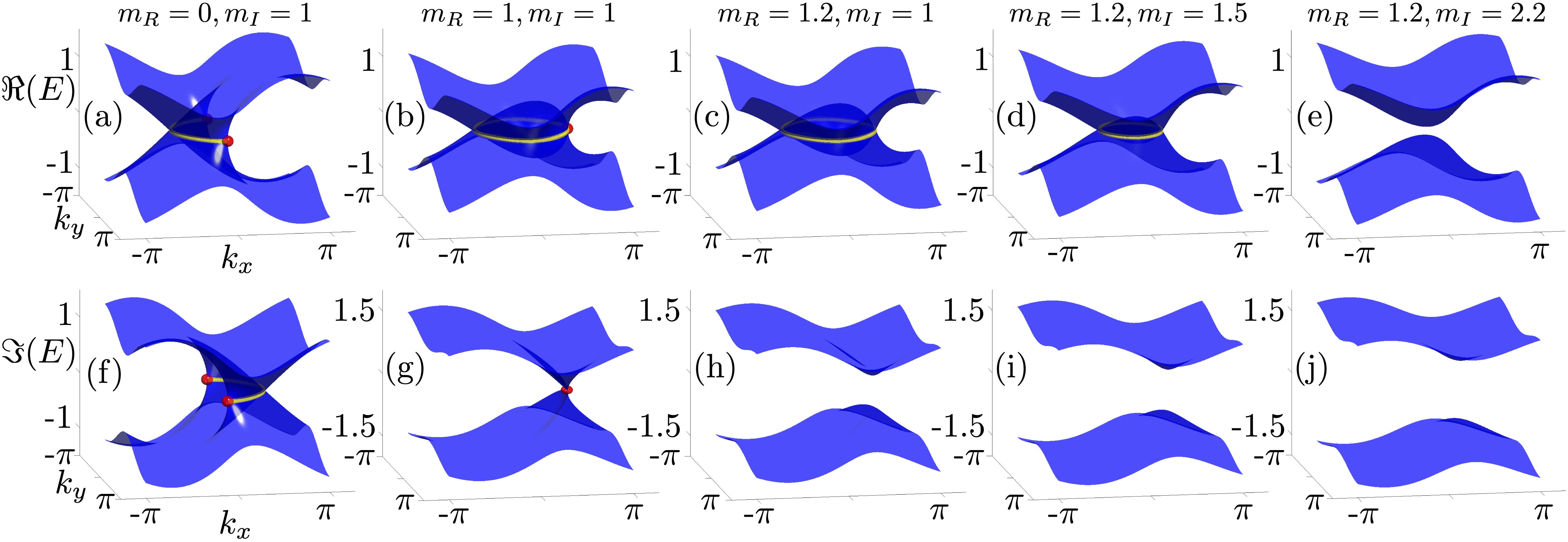}

\caption{Real and imaginary parts of the eigenvalues of the Hamiltonian given by Eq.~\eqref{eq:genEP2ann} in panels (a)-(e) and (f)-(j), respectively. 
The red points illustrate generic EP2s, with their concomitant bulk Fermi and i-Fermi arcs highlighted with the yellow lines in (a)-(e) and (f)-(j), respectively. 
By first increasing the magnitude of $m_R$ while keeping $m_I$ fixed, the EP2s eventually merge into a single EP2 [panels (b) and (g)], which is then immediately gapped out when $m_R$ is increased further [panels (c) and (h)]. 
While the (real) bulk Fermi arc is now forming a closed curve, the imaginary energy is completely gapped. 
This imaginary energy gap is kept when increasing the magnitude of $m_I$ [panels (i) and (j)].
In the real spectrum, varying $m_I$ instead shrinks the Fermi arc [panel (d)], to a point where it is fully removed [panel (e)].
Since this is occurring without passing through a gapless phase including an EP2, the Fermi arcs are not topological, explaining the trivial $\pi_d$-invariant predicted by the classification scheme.
\label{fig:GenEP2Ann} }
\end{figure*}

As have already been investigated for systems subject to non-local symmetries in Sec.~\ref{sec:NLsym}, the classification scheme relating non-Hermitian eigenvalue topology to Hermitian eigenvector topology through the resultant Hamiltonian does also say something about Abelian topology in the absence of EP$n$s.
Since this has already been resolved for systems subject to non-local symmetries, this section will again be devoted to generic systems and systems subject to local similarities.
Recalling that the generic (similarity-protected) EP$n$s themselves are classified by a $\pi_{2n-3}$ ($\pi_{n-2}$) invariant, the gapped topology is given by a $\pi_{2n-2}$ ($\pi_{n-1}$) invariant.
Quite surprisingly, this predicts that the eigenvalue topology of both generic and similarity-protected systems is trivial in the absence of EP$n$s. 
At first glance, this seems to contradict earlier results, such as the non-trivial braiding properties observed in gapped non-Hermitian systems. 
However, a non-trivial braid is often accompanied by a non-trivial Fermi arc, highlighting a connection between braids and the existence of bulk Fermi arcs.
The ability to form non-trivial braids in, for example, two-dimensional systems arises because the braid is non-trivial along a non-contractible circle on the torus, an aspect not captured by homotopy groups between spheres, but rather by those between the torus and the sphere~\cite{Wojcik2022}.  
Consequently, the trivial $\pi_d$-invariants derived using the resultant vector formalism do not prohibit the existence of topologically non-trivial bulk Fermi arcs that lie along any of the generating directions of the Brillouin torus in generic and similarity-protected non-Hermitian systems.  
What trivial $\pi_d$-invariants do say is rather that bulk Fermi arcs existing in the absence of EP$n$s that do not span any of the generating directions of the Brillouin torus, can be removed from the spectrum without necessitating the system to undergo a spectral phase transition involving an EP$n$.

To illustrate this claim, it is fruitful to resort to a low-dimensional example with few bands for intuition.
Consider a Hamiltonian given by
\begin{align}
    H(\bk) &= \begin{pmatrix} 0&1\\a_R(\bk)+i a_I(\bk)&0 \end{pmatrix}, \label{eq:genEP2ann}
    \\ 
    a_R(\bk) &= \sin k_x -m_R,
    \\
    a_I(\bk) &= m_I-\cos k_x-\cos k_y,
\end{align}
for some real-valued mass terms $m_R$ and $m_I$. 
Fig.~\ref{fig:GenEP2Ann} shows the corresponding real and imaginary spectrum. 
Upon varying $m_R$ and $m_I$, the two EP2s connected by an open bulk Fermi arc [Fig.~\ref{fig:GenEP2Ann} (a) and (f)], can annihilate in such a way that the spectrum is left with a closed bulk Fermi arc in the real spectrum [Fig.~\ref{fig:GenEP2Ann} (b)-(d)], while the imaginary spectrum only comprises the merged EP2 [Fig.~\ref{fig:GenEP2Ann} (g)] before it becomes completely gapped [Fig.~\ref{fig:GenEP2Ann} (h)-(j)]. 
As the magnitude of $m_I$ is increased, the bulk Fermi arc is shrinking [Fig.~\ref{fig:GenEP2Ann} (c)-(d)] to eventually be removed completely [Fig.~\ref{fig:GenEP2Ann} (e)]. 
If the Fermi arc was to be topological, such a process would require passing through a spectral phase including an EP2, but this is not the case here since the imaginary spectrum is kept fully gapped throughout this process.
It is possible since the generic and (local) similarity-protected non-Hermitian spectra are not constrained as the spectra of non-local symmetrical systems---there are no points acting as the TRIM points.
It should be noted, however, that the EP2s could equally well have been annihilated along any of the generating directions of the Brillouin torus.
This would result in one non-contractible, yet closed, Fermi arc spanning through all $k_x$ or $k_y$.
As mentioned above, such Fermi arcs are not classified by the strong topological invariants from resultant vector formalism, since they are not classified by maps between spheres, but rather by maps from the weak topological invariants given by the homotopy groups between a torus and a sphere~\cite{Wojcik2022}.
In other words, the resultant vector formalism captures local topology, i.e., topological properties of points in the Brillouin zone, rather than global topology of the full Brillouin zone.

\section{Concluding remarks} \label{sec:Con}
\subsection{Summary}
In this work, we have classified the Abelian eigenvalue topology of generic EP$n$s, as well as EP$n$s protected by local similarity relations and non-local symmetries, using a rigorous mathematical framework. 
Concretely, by encoding of EP$n$s in resultant vectors as singularities of vector fields on a Brillouin torus of appropriate dimensions, both generic and (local) similarity-protected EP$n$s are classified in terms of Mayer--Vietoris sequences. 
Furthermore, the exactness of these sequences leads to corresponding doubling theorems for EP$n$s. 
These theorems ensure that EP$n$s, similar to degeneracies in Hermitian matrices such as Weyl points, must appear in pairs within the Brillouin torus.
The Mayer--Vietoris sequences partly provide a geometric interpretation of the topological classification of EP$n$s, as it is related to the underlying classification of vector bundles, which is also worked out.
Apart from the interpretation in terms of Mayer--Vietoris sequences, the topological classification scheme of EP$n$s provided in Ref.~\cite{Yoshida2024} is furthermore extended to also include EP$n$s protected by self skew-similarity.
Through this classification, it becomes apparent that most generic and (local) similarity-protected EP$n$s are classified by generalized winding numbers corresponding to topological invariants of the Hermitian symmetry class AIII.
The exceptions are EP$n$s for even $n$ protected by pseudo-Hermiticity, which instead are classified by Chern numbers in symmetry class A when $n>2$, and a $\mathbb{Z}_2$-invariant in class D when $n=2$.
In contrast to the local similarities, non-local symmetries do not impact the codimension of EP$n$s, but instead induce additional eigenvalue topology.
This is shown to be intimately related to how EP$n$s are created and annihilated, as the symmetry-induced topology is sourced by topologically non-trivial bulk Fermi arcs connecting the EP$n$s
Upon annihilating the EP$n$s, the bulk Fermi arc itself can become topological meaning that it can only be removed from the spectrum by again passing through an EP$n$.
This predicts the existence of lone bulk Fermi arcs protected by a $\mathbb{Z}_2$-invariant, naturally dubbed $\mathbb{Z}_2$ Fermi arcs.
It is lastly shown that the general framework is not limited to point-like EP$n$s, but also leads to interesting conclusions for non-defective EP$n$s, EP$n$ of any dimension, and $n$-band systems without EP$n$s.
On a general level, the classification scheme describes Abelian eigenvalue topology of EP$n$s in non-Hermitian systems in terms of the tenfold way and Hermitian eigenvector topology.

\subsection{Discussion and outlook}
The general topological classification of EP$n$s, and the fundamental understanding of it, is of direct physical relevance in a large variety of modern research areas. 
Generic EP$n$s are realized in various systems with synthetic dimensions, such as Floquet systems~\cite{Zhou2023}. 
As for similarity-protected EP$n$s, pseudo-Hermitian matrices include the $\mathcal{PT}$-symmetric ones commonly used in optics. 
Here, $\mathcal{PT}$-symmetric matrices are used to describe reflectivity, or as effective Hamiltonian descriptions of photonic crystals with balanced gain and loss~\cite{Topphot,Lu2014,Ozdemir2019}. Self skew-similar systems include non-Hermitian Lieb lattices, which comprise yet another important extension to the non-Hermitian realm in photonics~\cite{Xiao,Vicencio2015,Mukherjee2015}. 
Further developing and increasing the theoretical understanding of the topological nature of EP$n$s will thus provide a stronger ground in all these, to mention only a handful, highly relevant research fields, which demonstrates the vast potential possessed by fundamental classification schemes.

Recent works have furthermore started to investigate what the physical consequences of having a non-orientable Brillouin zone, instead of the ordinary torus, are~\cite{fonesca2024}. 
On such manifolds, doubling theorems are modified and seem to take the form of mod 2 charge cancellations theorems (meaning that the indices/invariants locally assigned to singularities globally sum to an even number). 
Including these results in a more systematic and general classification scheme is desirable, and something to decipher in future works, both in the Hermitian and non-Hermitian realm.

The striking similarity between the classification of EP$n$s protected by pseudo-Hermitian similarity for even $n$, and Weyl semimetals, although connecting the eigenvalue topology of EP$n$s to the eigenvector topology of Weyl nodes, suggests a potential prediction of novel topological surface states in non-Hermitian systems. 
This stems from the existence of surface Fermi arcs in Weyl semimetals. 
One way of viewing this is from the respective Mayer--Vietoris sequences, which are equivalent for the two cases. 
For example, when describing Weyl semimetals, the Poincar\'e-dual sequence in (relative) homology groups is used to mathematically predict the existence of and to understand the topological properties of the surface Fermi arcs. 
Since the Mayer--Vietoris sequence of EP4s yields the very same Poincar\'e-dual sequence of homology groups as three-dimensional Weyl semimetals, it predicts the existence of something topologically equivalent to Fermi arcs in the space of resultants of the parent pseudo-Hermitian matrix.
Though being topological surface states in a resultant space, it is by no means guaranteed that these will correspond to surface states in the parameter space of the parent matrix. 
One should further take into account that the bulk-boundary correspondence is fundamentally different for non-Hermitian systems compared to Hermitian ones and that it further breaks down at the EPs due to the coalescence of eigenvectors~\cite{Kunst2018,Yao2018,Edvardsson2018}. 
Nevertheless, topological features in the space of resultants necessarily reflect some topological feature in the parent matrix, although the exact relation between these two might not be obvious. 
Revealing this connection would further increase the understanding and impact of Chern numbers in non-Hermitian topological band theory, and comprise a highly interesting and relevant question to answer in future works.

The classification of the Abelian topology of EP$n$s provides a strong theoretical foundation to be used to further develop this field. 
Naturally, unraveling non-Abelian topology corrections physically stemming from adding more bands, or, mathematically, from placing an existing $n\times n$ Jordan block into a larger square matrix, is one possible direction. 
A particularly interesting phenomenon directly linked to this is the appearance of different kinds of EPs of a particular order in the same matrix.
A recent work has initiated such a study, albeit still within the Abelian regime, by classifying ``Hopf EPs''~\cite{Yoshida2025}.
These are exactly EPs whose codimension is generically higher than what is expected from the conventional argument based on counting constraints. 
Additionally, the breakdown of the Abelian classification based on the resultant winding number is also emphasized through the introduction of ``fake EPs''---points in momentum space where the resultant winding number is non-trivial despite the absence of an EP.
This phenomena arise exclusively when the size of the matrix is larger than the degree of the degeneracy, showing the importance of extending the existing Abelian classification schemes to the non-Abelian realm in future works.

Despite being developed to unravel the topological properties of EP$n$s in non-Hermitian systems, the classification schemes may find profound implications also in Hermitian band structures.
Arguably, the most prominent example is given by EP$n$s protected by non-local symmetries and the intricate relation between bulk Fermi arcs in non-Hermitian systems, and Dirac strings in Hermitian systems.
This serves as an additional physical motivation to further unravel the underlying vector bundle structure related to $\mathcal{T}$-symmetric systems in dimensions beyond four.
This would not only further deepen the understanding of the symmetry-induced topology related to EP$n$s, but also provide a more fundamental description of the generalizations of FKM-invariants and their corresponding monopoles in Hermitian systems.
Consequently, this provides a recipe towards explicit constructions of higher-dimensional topological insulators models with non-trivial $\mathbb{Z}_2$-topology using simpler non-Hermitian models.
The resultant Hamiltonians corresponding to the non-Hermitian toy models studied in Sec.~\ref{sec:NLPhys} comprise concrete such examples.

Finally, it is again worth emphasizing the striking connection between abstract mathematical concepts and physical properties of direct experimental relevance. 
Bridging these seemingly distant fields is done through the theoretical framework of topological band theory and, more concretely, here in terms of Abelian classification schemes of the topological properties of EP$n$s. 
Extending this venue further, and simultaneously elucidating its topological and physical consequences, comprises a highly relevant step towards unraveling the full richness of non-Hermitian topological band theory.

\section*{Acknowledgements}
The authors thank Tsuneya Yoshida, Emil J. Bergholtz, Lukas K\"onig, Thijs Douwes, Flore K. Kunst, and Anton Montag for related collaborations and discussions. 
M.S. is supported by the Swedish Research Council (VR) under grant NO. 2024-00272.
L.R. is supported by the Knut and Alice Wallenberg Foundation under Grant No. 2017.0157.

\end{document}